\documentclass{revtex4}

\usepackage{amsthm,amssymb,amsmath}				
\usepackage{graphicx,comment}
\usepackage{float}   
\usepackage{xcolor}

\begin{document}

\title{PDM KG-Coulombic particles in cosmic string rainbow gravity spacetime
and a uniform magnetic field }
\author{Omar Mustafa}
\email{omar.mustafa@emu.edu.tr}
\affiliation{Department of Physics, Eastern Mediterranean University, G. Magusa, north
Cyprus, Mersin 10 - Turkey.}

\begin{abstract}
\textbf{Abstract:}\ Klein-Gordon (KG) particles in cosmic string rainbow
gravity spacetime and a uniform magnetic field are studied in the context of
the so called, metaphorically speaking, position-dependent mass (PDM)
settings. We show that the corresponding KG-equation collapses into a
two-dimensional radial Schr\"{o}dinger-Coulomb like model. The exact
textbook solution of which is used to find the energies and wave
functions of KG-Coulombic particles (both constant mass and PDM ones). In so
doing, we consider, with $y=E/E_{P}$, four pairs of rainbow functions: (a) $%
g_{_{0}}\left( y\right) =1$, $g_{_{1}}\left( y\right) =\sqrt{1-\epsilon y^{2}%
}$, (b) $g_{_{0}}\left( y\right) =1$, $g_{_{1}}\left( y\right) =\sqrt{%
1-\epsilon y}$, (c) $g_{_{0}}\left( y\right) =g_{_{1}}\left( y\right)
=\left( 1-\epsilon y\right) ^{-1}$, and (d) $g_{_{0}}( y) =(e^{\epsilon y}-1) /\epsilon y$, $g_{_{1}}\left( y\right) =1$. Interestingly, we observe that the first pair in (a) introduces the Planck energy $E_{p}$ as a maximum possible  KG-particle/antiparticle energy value.


\end{abstract}

\maketitle

\section{Introduction}

Rainbow gravity (RG) has attracted research attention over the years \cite%
{R1,R2,R3,R4,R5} as a semi-classical extension of the deformed/doubly
special relativity into general relativity (GR). It suggests that the energy
of the probe particles affects the spacetime background, at the ultra-high
energy regime, and the spacetime metric becomes energy-dependent \cite%
{R5,R6,R7,R8,R9,R10,R11,R12,R13,R14,R15,R16,R17}. That is, a cosmic string
spacetime metric (in the natural units $c=\hbar =G=1$) 
\begin{equation}
ds^{2}=-dt^{2}+dr^{2}+\alpha ^{2}\,r^{2}d\varphi ^{2}+dz^{2},  \label{e1}
\end{equation}%
would, under RG, take an energy-dependent form%
\begin{equation}
ds^{2}=-\frac{1}{g_{_{0}}\left( y\right) ^{2}}dt^{2}+\frac{1}{g_{_{1}}\left(
y\right) ^{2}}\left( dr^{2}+\alpha ^{2}\,r^{2}d\varphi ^{2}+dz^{2}\right)
;\;y=E/E_{p},  \label{e4}
\end{equation}
where $\alpha $ is a constant related to the deficit angle of the conical
spacetime and is defined as $\alpha =1-4G\mu $, $G$ is the Newton's
constant, $\mu $ is the linear mass density of the cosmic string so that $%
\alpha <1,E$ is the energy of the probe particle, and $E_{p}=\sqrt{\hbar
c^{5}/G}$ is the Planck energy. Here, the signature of the line elements (\ref{e1}) and (\ref{e4}) is $\left( -,+,+,+\right) $. Moreover, the
corresponding metric tensor $g_{\mu \nu }$ reads%
\begin{equation}
g_{\mu \nu }=diag\left( -\frac{1}{g_{_{0}}\left( y\right) ^{2}},\frac{1}{%
g_{_{1}}\left( y\right) ^{2}},\frac{\alpha ^{2}\,r^{2}}{g_{_{1}}\left(
y\right) ^{2}},\frac{1}{g_{_{1}}\left( y\right) ^{2}}\right) ;\;\mu ,\nu
=t,r,\varphi ,z,  \label{e5}
\end{equation}%
and 
\begin{equation}
\det ( g_{\mu \nu }) =-\frac{\alpha ^{2}\,r^{2}}{g_{_{0}}(
y) ^{2}g_{_{1}}( y) ^{6}}\Longrightarrow g^{\mu \nu
}=diag( -g_{_{0}}( y) ^{2},g_{_{1}}( y) ^{2},%
\frac{g_{_{1}}( y) ^{2}}{\alpha ^{2}\,r^{2}},g_{_{1}}(
y) ^{2}) .  \label{e6}
\end{equation}%
where $g_{_{0}}\left( y\right) $, $g_{_{1}}\left( y\right) $ are the rainbow functions.

The Planck energy $E_{p}$, in the RG model, is considered to represent a
threshold separating classical from quantum mechanical descriptions to
introduce itself as yet another invariant energy scale alongside the speed
of light. Consequently, rainbow gravity justifies the modification of the
relativistic energy-momentum dispersion relation into%
\begin{equation}
E^{2}g_{_{0}}\left( y\right) ^{2}-p^{2}c^{2}g_{_{1}}\left( y\right)
^{2}=m^{2}c^{4};\;0\leq \left( y=E/E_{p}\right) \leq 1,  \label{e2}
\end{equation}%
where $mc^{2}$ is its rest mass energy. Such a modification is significant
in the ultraviolet limit and is constrained to reproduce the standard GR
dispersion relation in the infrared limit so that 
\begin{equation}
\lim\limits_{y\rightarrow 0}g_{_{k}}\left( y\right) =1;\;k=0,1.  \label{e3}
\end{equation}%
The effects of such modifications could be observed, for example, in the
tests of thresholds for ultra high-energy cosmic rays \cite{R6,R13,R14,R15},
TeV photons \cite{R16}, gamma-ray bursts \cite{R6}, nuclear physics
experiments \cite{R17}. Rainbow gravity settings have motivated interesting
recent studies on the associated quantum gravity effects. Such studies
include, for example, the thermodynamical properties of black holes \cite%
{R18,R19,R20,R21,R211}, the dynamical stability conditions of neutron stars 
\cite{R22}, thermodynamic stability of modified black holes \cite{R221},
charged black holes in massive RG \cite{R222}, on geometrical thermodynamics
and heat engine of black holes in RG \cite{R223}, on RG and f(R) theories 
\cite{R224}, the initial singularity problem for closed rainbow cosmology 
\cite{R23}, the black hole entropy \cite{R24}, the removal of the
singularity of the early universe \cite{R25}, the Casimir effect in the
rainbow Einstein's universe \cite{R8}, massive scalar field in RG
Schwarzschild metric \cite{R26}, five-dimensional Yang--Mills black holes in
massive RG \cite{R27}, etc.

On the other hand, the dynamics of Klein-Gordon (KG) particles (i.e., spin-0
mesons), Dirac particles (spin-1/2 fermionic particles), and
Duffen-Kemmer-Peatiau (DKP) particles (spin-1 particles like bosons and
photons) in different spacetime backgrounds in rainbow gravity are studied.
For example, in a cosmic string spacetime background in rainbow gravity,
Bezzerra et al. \cite{R8,R81} have studied Landau levels via Schr\"{o}dinger and
KG equations, Bakke and Mota \cite{R28} have studies the Dirac oscillator,
they have also studied the Aharonov-Bohm effect \cite{R29}. Hosseinpour et
al. \cite{R5} have studied the DKP-particles, Sogut et al. \cite{R11} have
studied the quantum dynamics of photon, Kangal et al. \cite{R12} have
studied KG-particles in a topologically trivial G\"{o}del-type spacetime in
rainbow gravity, and very recently KG-oscillators in cosmic string rainbow
gravity spacetime in a non-uniform magnetic field are studied by Mustafa 
\cite{R291} (without and with the position-dependent mass (PDM) settings).
In the current proposal, however, we extend such studies and consider PDM
KG-Coulombic particles in cosmic string rainbow gravity spacetime and a
uniform magnetic field.

One should be reminded, nevertheless, that PDM is a metaphoric notion that
emerges as a manifestation of coordinate transformation/deformation that
renders the mass to become effectively position-dependent \cite%
{R30,R31,R32,R33,R34,R35,R36,R37}. PDM concept has been introduced \ in the
study PDM KG-oscillators in cosmic string spacetime within Kaluza-Klein
theory \cite{R38}, in (2+1)-dimensional G\"{u}rses spacetime backgrounds 
\cite{R39}, and in Minkowski spacetime with space-like dislocation \cite{R40}%
. Basically, for the PDM von Roos Schr\"{o}dinger Hamiltonian \cite{R30}, it
has been shown (c.f., e.g., \cite{R31,R32,R33}) that the PDM momentum
operator takes the form $\hat{p}_{j}( \mathbf{r}) =-i[
\partial _{j}-\partial _{j}f( \mathbf{r}) /4f( \mathbf{r}) ] \,;\;j=1,2,3,$ where $f\left( \mathbf{r}\right) $ is a
positive-valued dimensionless scalar multiplier. For more details on this
issue the reader may refer to \cite{R31,R33,R38,R39,R40,R401,R402}. This assumption
would, in turn, allow one to cast the PDM von Roos kinetic energy operator
(using $\hbar =2m=1$ units in the von Roos Hamiltonian) as $\hat{T}( 
\mathbf{r})\psi ( \mathbf{r}) =-f( \mathbf{r}) ^{-1/4}( \mathbf{\nabla \,}f( \mathbf{r}) ^{-1/2}) \cdot ( \mathbf{\nabla \,}%
f( \mathbf{r}) ^{-1/4}\psi ( \mathbf{r}) ) $
(known in the literature as Mustafa-Mazharimousavi's PDM kinetic energy
operator \cite{R32}). Which suggests that the momentum operator for constant
mass setting, $\hat{p}_{j}-i\,\partial _{j}$, should be replaced by the PDM
operator $\hat{p}_{j}\left( \mathbf{r}\right) $ for PDM settings. We shall
use such PDM recipe in the current study of PDM KG-Coulombic particles in
cosmic string rainbow gravity spacetime and a uniform magnetic field. We shall
be interested in three pairs of rainbow functions: (a) $g_{_{0}}\left(
y\right) =1$, $g_{_{1}}\left( y\right) =\sqrt{1-\epsilon y^{2}}$, and $%
g_{_{0}}\left( y\right) =1$, $g_{_{1}}\left( y\right) =\sqrt{1-\epsilon y}$,
which belong to the set of rainbow functions $g_{_{0}}\left( y\right) =1$, $%
g_{_{1}}\left( y\right) =\sqrt{1-\epsilon y^{n}}$ (where $\epsilon $ is a
dimensionless constant of order unity) used to describe the geometry of
spacetime in loop quantum gravity \cite{R41,R42}, (b) $g_{_{0}}\left(
y\right) =g_{_{1}}\left( y\right) =\left( 1-\epsilon y\right) ^{-1}$, a
suitable set used to resolve the horizon problem \cite{R13,R43}, and (c) $%
g_{_{0}}( y) =( e^{\epsilon y}-1) /\epsilon y$ and $%
g_{_{1}}\left( y\right) =1$, which are obtained from the spectra of gamma-ray bursts at cosmological distances \cite{R6}.

The organization of our paper is in order. We discuss, in section 2, PDM
KG-particles in the cosmic string rainbow gravity spacetime (\ref{e4}) and a
uniform magnetic field. We show that the corresponding KG-equation collapses
into the two-dimensional radial Schr\"{o}dinger-Coulomb equation. In section
3, we discuss the RG effect (using the above mention rainbow functions sets)
on the spectroscopic structure of KG-Coulombic constant mass particles. We
discuss, in section 4, the effects of rainbow gravity as well as PDM on the
energy levels of a PDM KG-Coulombic particle. Our concluding remarks are
given in section 5.

\section{PDM KG-particles in cosmic string rainbow gravity spacetime and a uniform magnetic field}
In the cosmic string rainbow gravity spacetime background (\ref{e4}), a
KG-particle of charge $e$ in a 4-vector potential $A_{\mu }$ is described
(in $c=\hbar =G=1$ units) by the KG-equation%
\begin{equation}
\frac{1}{\sqrt{-g}}D_{\mu }\left( \sqrt{-g}g^{\mu \nu }D_{\nu }\Psi \right)
=m^{2}\Psi ,  \label{e7}
\end{equation}%
where $D_{\mu }$ is the gauge-covariant derivative given by $D_{\mu
}=\partial _{\mu }-ieA_{\mu }$, and $m$ is the rest mass energy of the
KG-particle. At this point, we may also include position-dependent mass
(PDM) settings (a metaphoric description of deformed coordinates and
inherited from the von Roos Hamiltonian \cite{R30} ) using the PDM-momentum
operator $\hat{p}_{j}( \mathbf{r}) =-i[ \partial
_{j}-\partial _{j}\,f( \mathbf{r}) /4\,f( \mathbf{r})] $ \cite{R31,R32,R33,R38,R39,R40,R401,R402}. In this case, $D_{\mu
}\longrightarrow \tilde{D}_{\mu }=D_{\mu }+\mathcal{F}_{\mu }=\partial _{\mu
}+\mathcal{F}_{\mu }-ieA_{\mu }$, where $\mathcal{F}_{\mu }=( 0,\mathcal{F}_{r},0,0) $, $\mathcal{F}_{r}=f^{\prime }\left( r\right)
/4\,f\left( r\right) $ and $f\left( \mathbf{r}\right) =f\left( r\right) $ is
only radially dependent. One should notice that a KG-oscillator is obtained
using $f( r) =\exp ( 2\beta r^{2}) $, where $f\left(
r\right) $ is a positive dimensionless scalar multiplier. Under such new
structure our KG-equation (\ref{e7}) now describes PDM KG-particles in the
cosmic string rainbow gravity spacetime and reads%
\begin{equation}
\frac{1}{\sqrt{-g}}\tilde{D}_{\mu }\left( \sqrt{-g}g^{\mu \nu }\tilde{D}%
_{\nu }\right) \Psi =m^{2}\Psi \Longrightarrow \frac{1}{\sqrt{-g}}\left(
D_{\mu }+\mathcal{F}_{\mu }\right) \sqrt{-g}g^{\mu \nu }\left( D_{\nu }-%
\mathcal{F}_{\nu }\right) \Psi =m^{2}\Psi .  \label{e8}
\end{equation}%
Which, in a straightforward manner, yields%
\begin{equation}
\left\{ -g_{_{0}}\left( y\right) ^{2}\partial _{t}^{2}+g_{_{1}}\left(
y\right) ^{2}\left[ \partial _{r}^{2}+\frac{1}{r}\partial _{r}-M\left(
r\right) +\frac{1}{\alpha ^{2}\,r^{2}}\left( \partial _{\varphi
}-ieA_{\varphi }\right) ^{2}+\partial _{z}^{2}\right] \right\} \Psi \left(
t,r,\varphi ,z\right) =m^{2}\Psi \left( t,r,\varphi ,z\right) ,  \label{e9}
\end{equation}%
where%
\begin{equation}
M\left( r\right) =\mathcal{F}_{r}^{\prime }+\frac{\mathcal{F}_{r}}{r}+%
\mathcal{F}_{r}^{2}=-\frac{3}{16}\left( \frac{f^{\prime }\left( r\right) }{%
f\left( r\right) }\right) ^{2}+\frac{f^{\prime }\left( r\right) }{4rf\left(
r\right) }+\frac{f^{\prime \prime }\left( r\right) }{4f\left( r\right) }
\label{e10}
\end{equation}%
We now use the substitution%
\begin{equation}
\Psi \left( t,r,\varphi ,z\right) =\exp \left( i\left[ \ell \varphi
+k_{z}z-Et\right] \right) \psi \left( r\right) ,  \label{e11}
\end{equation}%
in Eq. (\ref{e9}) to obtain%
\begin{equation}
\left\{ \tilde{E}^{2}+g_{_{1}}\left( y\right) ^{2}\left[ \partial _{r}^{2}+%
\frac{1}{r}\partial _{r}-M\left( r\right) -\frac{\left( \ell -eA_{\varphi
}\right) ^{2}}{\alpha ^{2}\,r^{2}}\right] \right\} \psi \left( r\right) =0,
\label{e12}
\end{equation}%
where 
\begin{equation}
\tilde{E}^{2}=g_{_{0}}\left( y\right) ^{2}E^{2}-g_{_{1}}\left( y\right)
^{2}k_{z}^{2}-m^{2}  \label{e13}
\end{equation}

In what follows we shall consider $A_{\varphi }=\frac{1}{2}B_{\circ }r$,
which in turn yields a non-uniform magnetic field $\mathbf{B}=\mathbf{\nabla 
}\times \mathbf{A}=B_{\circ }\,\hat{z}$. Consequently, Eq.(\ref{e12}) becomes%
\begin{equation}
\left\{ \lambda +\partial _{r}^{2}+\frac{1}{r}\partial _{r}-M\left( r\right)
-\frac{\tilde{\ell}^{2}}{r^{2}}+\frac{\tilde{\ell}\,\tilde{B}}{r}\right\}
\psi \left( r\right) =0,  \label{e14}
\end{equation}%
where%
\begin{equation}
\lambda =\frac{g_{_{0}}\left( y\right) ^{2}E^{2}-g_{_{1}}\left( y\right)
^{2}\left( k_{z}^{2}+\frac{\tilde{B}^{2}}{4}\right) -m^{2}}{g_{_{1}}\left(
y\right) ^{2}},\;\tilde{\ell}=\frac{\ell }{\alpha },\;\tilde{B}=\frac{%
eB_{\circ }}{\alpha }.  \label{e15}
\end{equation}%
Moreover, with $\psi \left( r\right) =R\left( r\right) /\sqrt{r}$ we obtain 
\begin{equation}
\left\{ \partial _{r}^{2}-\frac{\left( \tilde{\ell}^{2}-1/4\right) }{r^{2}}%
-M\left( r\right) +\frac{\tilde{\ell}\,\tilde{B}}{r}+\lambda \right\}
R\left( r\right) =0.  \label{e16}
\end{equation}%
Under such spacetime and magnetic field structures, we shall consider two
types of KG-particles: constant mass and PDM ones.

\section{Constant mass KG-particles in cosmic string rainbow gravity spacetime and a uniform magnetic field }
It is convenient to discuss the KG-particles with a standard constant mass,
i.e., $f\left( r\right) =1\Longleftrightarrow $ $M\left( r\right) =0$, so
that Eq.(\ref{e16}) reduces into the two-dimensional Schr\"{o}%
dinger-oscillator form%
\begin{equation}
\left\{ \partial _{r}^{2}-\frac{\left( \tilde{\ell}^{2}-1/4\right) }{r^{2}}+%
\frac{\tilde{\ell}\,\tilde{B}}{r}+\lambda \right\} R\left( r\right) =0.
\label{e17}
\end{equation}%
Which obviously admits exact solution in the form of hypergeometric function
so that%
\begin{equation}
R\left( r\right) \sim \,\left( 2i\sqrt{\lambda }r\right) ^{|\tilde{\ell}%
|+1/2}\exp \left( -i\sqrt{\lambda }r\right) \,_{1}F_{1}\left( \frac{1}{2}+%
\tilde{\ell}|-\frac{\tilde{\ell}\,\tilde{B}}{2i\sqrt{\lambda }},1+2|\tilde{%
\ell}|,2i\sqrt{\lambda }r\right) .  \label{e18}
\end{equation}%
However, to secure finiteness and square integrability we need to terminate
the hypergeometric function into a polynomial of degree $n_{r}\geq 0$ so
that the condition%
\begin{equation}
\frac{1}{2}+\tilde{\ell}|-\frac{\tilde{\ell}\,\tilde{B}}{2i\sqrt{\lambda }}%
=-n_{r}.  \label{e19}
\end{equation}%
is satisfied. This would in turn imply that%
\begin{equation}
i\sqrt{\lambda }=\frac{\tilde{\ell}\,\tilde{B}}{2\tilde{n}};\;\tilde{n}%
=n_{r}+|\tilde{\ell}|+\frac{1}{2}\Rightarrow \lambda _{n_{r},\ell }=-\frac{%
\tilde{\ell}\,^{2}\tilde{B}^{2}}{4\tilde{n}^{2}},  \label{e20}
\end{equation}%
and%
\begin{equation}
\psi \left( r\right) =\frac{R\left( r\right) }{\sqrt{r}}=\mathcal{N}\,r^{|%
\tilde{\ell}|}\exp \left( -\frac{|\tilde{\ell}\,\tilde{B}|}{2\tilde{n}}%
\,r\right) \,_{1}F_{1}\left( -n_{r},1+2|\tilde{\ell}|,\frac{|\tilde{\ell}\,%
\tilde{B}|}{\tilde{n}}r\right) .  \label{e201}
\end{equation}%
Consequently, Eq.(\ref{e15}) would read%
\begin{equation}
g_{_{0}}\left( y\right) ^{2}E^{2}-m^{2}=g_{_{1}}\left( y\right)
^{2}\,K_{n_{r},\ell };\;K_{n_{r},\ell }=\frac{\tilde{B}^{2}}{4}\left[ 1-%
\frac{\tilde{\ell}^{2}}{\left( n_{r}+|\tilde{\ell}|+\frac{1}{2}\right) ^{2}}%
\right] +k_{z}^{2}.  \label{e21}
\end{equation}%
At this point, it is convenient to mention that the choice of $i\sqrt{%
\lambda }=\tilde{\ell}\,\tilde{B}/2\tilde{n}\geq 0\Rightarrow i\sqrt{\lambda 
}=|\tilde{\ell}\,\tilde{B}|/2\tilde{n}$ is manifested by the requirement of
finiteness and square integrability of $\psi \left( r\right) $ as $%
r\rightarrow \infty $. Interestingly, moreover, we notice that all $S$-states (i.e., $\ell =0$ states) are degenerate  with each other (positive with positive and negative with negative states) and have the same $%
K_{n_{r},0}$ value 
\begin{equation}
K_{0,0}=K_{1,0}=\cdots =K_{n_{r},0}=\tilde{B}^{2}/4+k_{z}^{2}  \label{e21-1}
\end{equation}
as suggested by Eq.(\ref{e21}). This is a consequence of the cosmic string
spacetime (i.e., at $\lim\limits_{\epsilon \rightarrow 0}g_{_{0}}\left(
y\right) =\lim\limits_{\epsilon \rightarrow 0}g_{_{1}}\left( y\right) =1$)
and has nothings to do with rainbow gravity. Moreover, for a given $n_{r}$
and $\ell $ we have $K_{n_{r},|\ell |}=K_{n_{r},-|\ell |}$, which indicates
that we shall have eminent degeneracies associated with the magnetic quantum
number $\ell =\pm |\ell |\neq 0$ for each radial quantum number $n_{r}$.
These effects are going to be reflected on the spectroscopic structure of
constant mass KG-particles in cosmic string rainbow gravity spacetime, under
different sets of rainbow functions.

To observe the rainbow gravity effects on such constant mass KG-particles we
now consider different rainbow functions.

\subsection{The set of rainbow functions $g_{_{0}}\left( y\right) =1$, $%
g_{_{1}}\left( y\right) =\sqrt{1-\epsilon y^{n}}$}

We start with the set $g_{_{0}}\left( y\right) =1$ and $g_{_{1}}\left(
y\right) =\sqrt{1-\epsilon y^{2}}$ (i.e., $n=2$). Under such rainbow
functions structure the energy levels of (\ref{e21}) are given by 
\begin{equation}
E^{2}-m^{2}=\left( 1-\epsilon \frac{E^{2}}{E_{p}^{2}}\right) K_{n_{r},\ell
}\Longrightarrow E_{n_{r},\ell }=\pm \sqrt{\frac{K_{n_{r},\ell }+m^{2}}{%
1+\delta \,K_{n_{r},\ell }}};\;\delta =\frac{\epsilon }{E_{p}^{2}}.
\label{e24}
\end{equation}%
We plot, in Figure 1(a), the corresponding energies against $\delta
=\epsilon /E_{p}^{2}$. We observe, for a given radial quantum number $n_{r}$%
, eminent clustering of positive/negative energy levels as $\delta $ grows
up from zero (i.e., the cosmic string spacetime limit). In Figure 1(b),
moreover, we plot the energies against $\left\vert eB_{\circ }\right\vert $.
It is obvious that as $\left\vert eB_{\circ }\right\vert \rightarrow 0$ the
energy levels converge to the values $E_{n_{r},\ell }\sim \pm \sqrt{\left(
k_{z}^{2}+m^{2}\right) /\left( 1+\delta k_{z}^{2}\right) }=\sqrt{2/1.1}$ for 
$\delta =0.1$, and $m=k_{z}=1$ value used here. That is, at this limit positive/negative energy states emerge from the same
positive/negative values irrespective of the values of the radial and the
magnetic quantum numbers $n_{r}$ and $\ell $. 
On the other hand, as $%
\left\vert eB_{\circ }\right\vert >>1$, the energy levels cluster and merge
into $E_{n_{r},\ell }\sim \pm \sqrt{1/\delta }$ (this is observed in Figure
1(b), i.e., as $\left\vert eB_{\circ }\right\vert >>1$ the energies $%
E_{n_{r},\ell }\sim \pm \sqrt{1/\delta }=\pm \sqrt{10}$ for $\delta =0.1$).
 We may, therefore, conclude that under such rainbow functions structure the energy levels are destined
to be within the range $\sqrt{( k_{z}^{2}+m^{2}) /( 1+\delta
k_{z}^{2}) }\leq |E_{n_{r},\ell }|\,\leq \sqrt{1/\delta }=E_{p}/\sqrt{%
\epsilon }$.%
\begin{figure}[!ht]  
\centering
\includegraphics[width=0.35\textwidth]{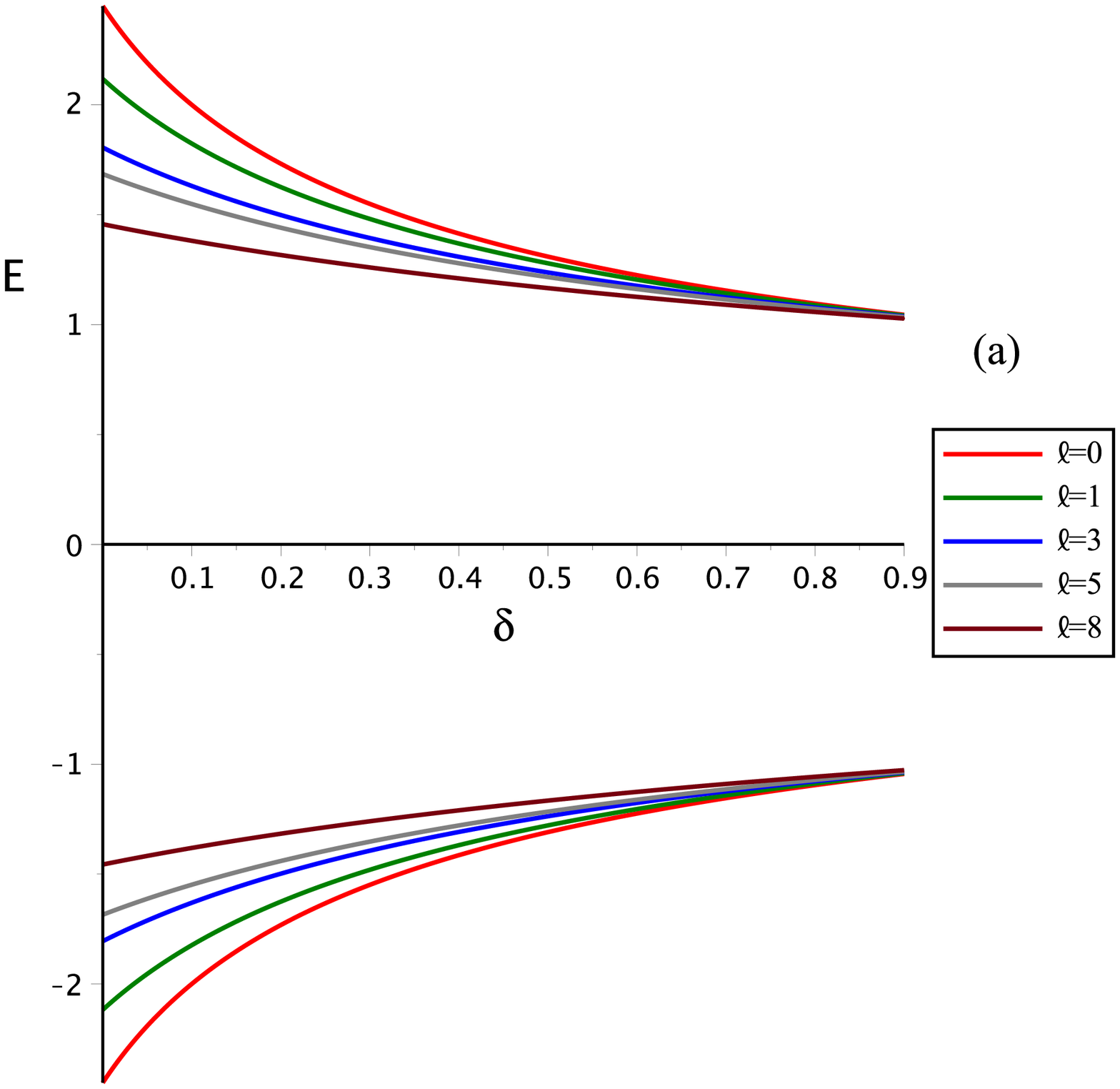}
\includegraphics[width=0.35\textwidth]{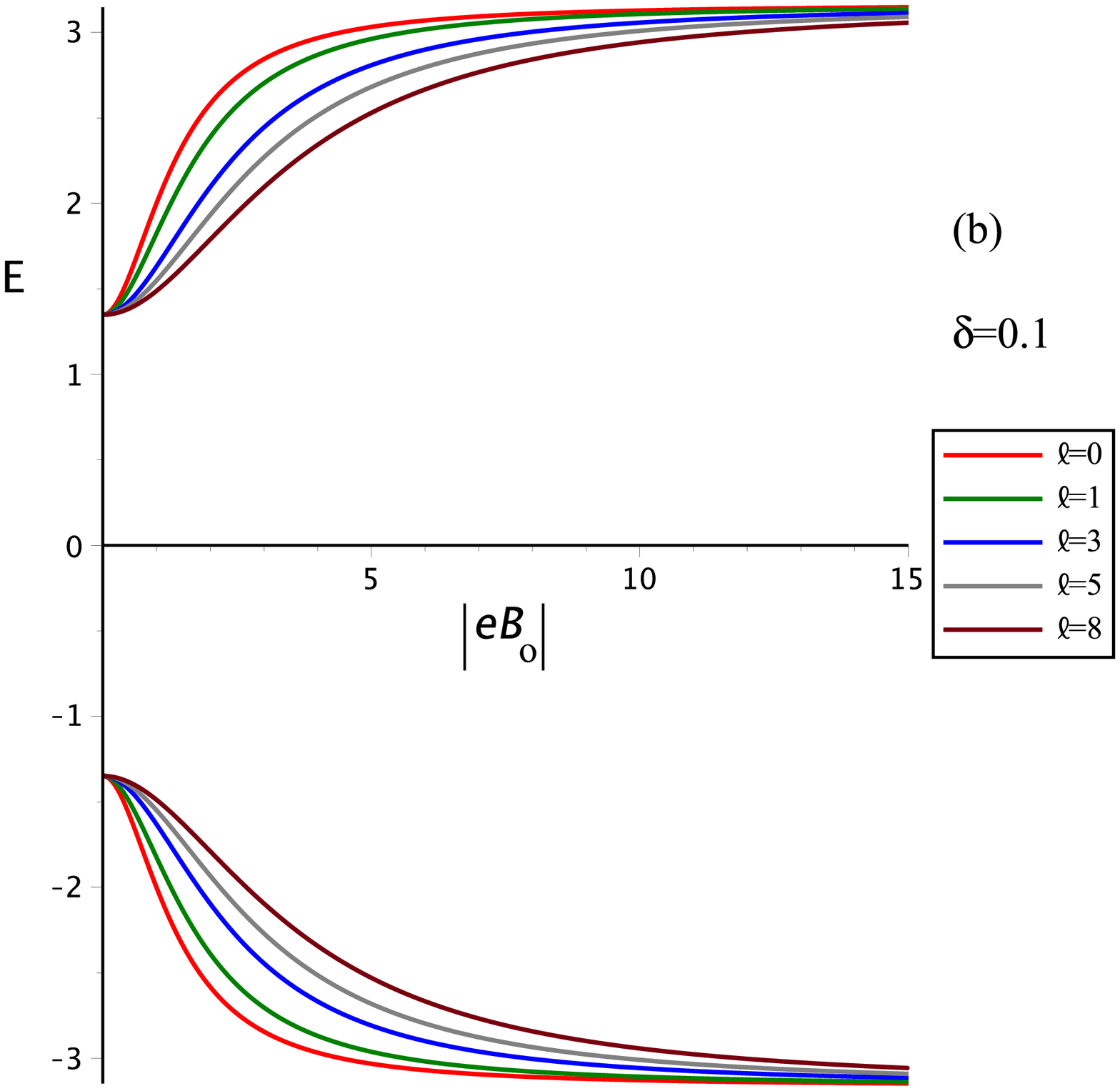} 
\caption{\small 
{ The energy levels of (\ref{e24}), using $\alpha =1/4$, $%
m=k_{z}=1$, so that (a) shows $E$ against $\delta =\epsilon /E_{p}^{2}$ for $%
|eB_{\circ }|=1$, $n_{r}=2$, $\ell =0,1,3,5,8$, and (b) shows $E$ against $%
|eB_{\circ }|$ for $\delta =0.1$, $n_{r}=2$, $\ell =0,1,3,5,8$.}}
\label{fig1}
\end{figure}%
\begin{figure}[!ht]  
\centering
\includegraphics[width=0.35\textwidth]{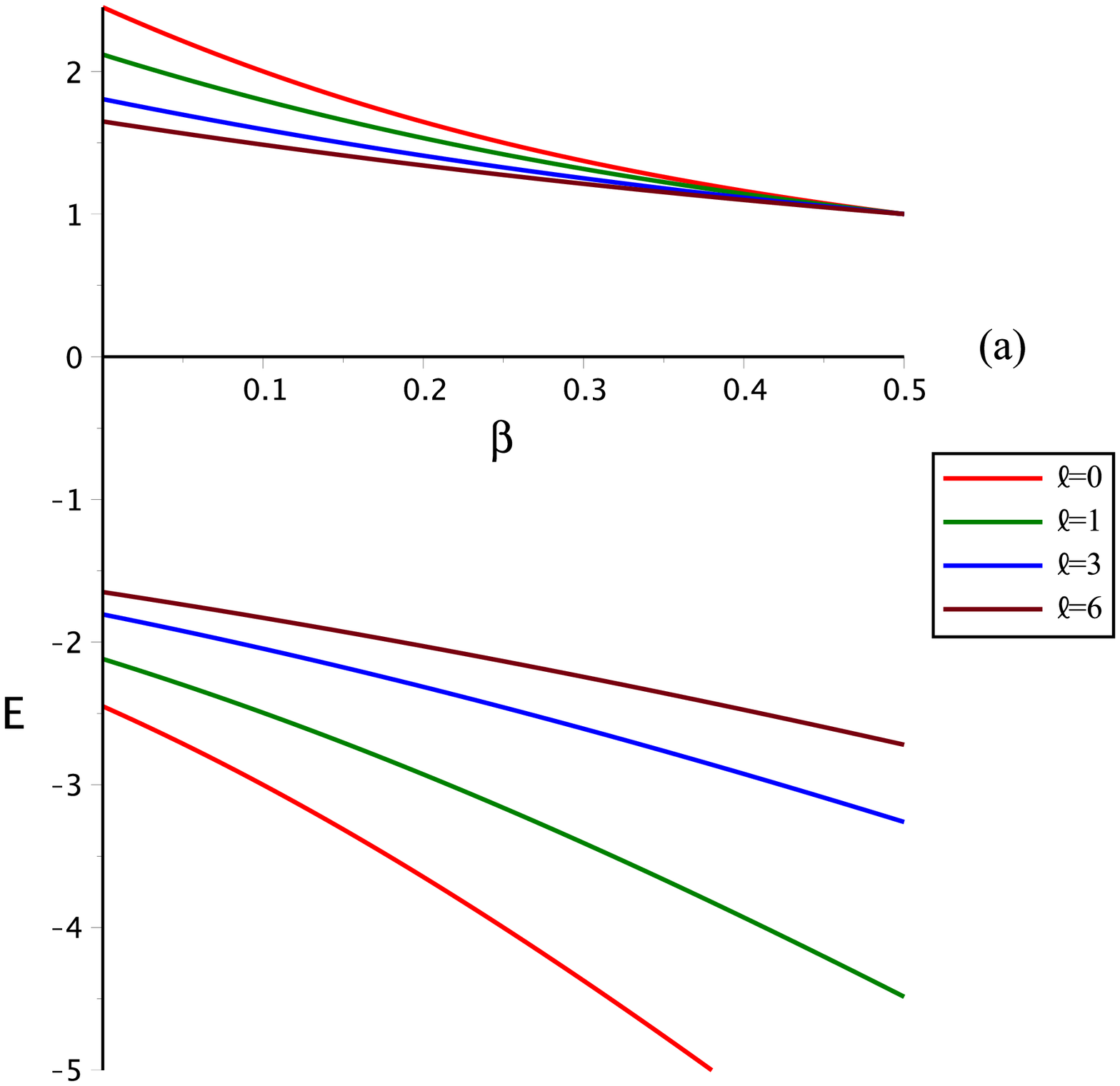}
\includegraphics[width=0.35\textwidth]{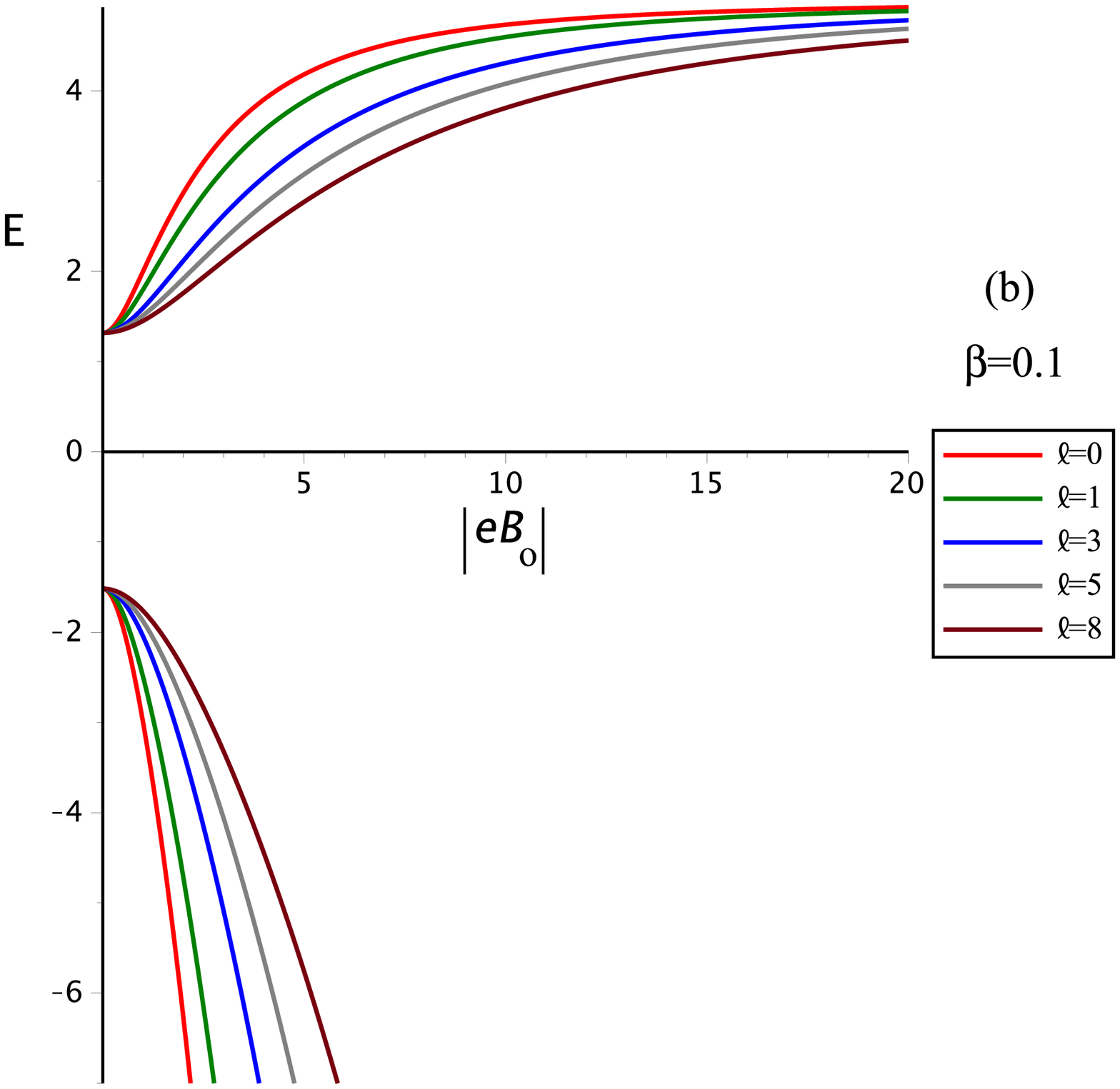} 
\caption{\small 
{ The energy levels of (\ref{e25}), using $\alpha =1/4$, $%
m=k_{z}=1$, so that (a) shows $E$ against $\beta =\epsilon /2E_{p}$ for $%
|eB_{\circ }|=1$, $n_{r}=2$, $\ell =0,1,3,6$, and (b) shows $E$ against $%
|eB_{\circ }|$ for $\beta =0.1$, $n_{r}=2$, $\ell =0,1,3,5,8$.}}
\label{fig2}
\end{figure}%

Now we consider the set $g_{_{0}}\left( y\right) =1$ and $g_{_{1}}\left(
y\right) =\sqrt{1-\epsilon y}$ (i.e., $n=1$) in Eq.(\ref{e21}) to obtain%
\begin{equation}
E^{2}-m^{2}=\left( 1-\epsilon \frac{E}{E_{p}}\right) K_{n_{r},\ell
}\Longrightarrow E_{n_{r},\ell }=-\beta K_{n_{r},\ell }\pm \sqrt{\beta
^{2}K_{n_{r},\ell }^{2}+K_{n_{r},\ell }+m^{2}};\;\beta =\frac{\epsilon }{%
2E_{p}}.  \label{e25}
\end{equation}%
In Figures 2(a) and 2(b), we plot the energy levels against $\beta =\epsilon
/2E_{p}$ and $\left\vert eB_{\circ }\right\vert $, respectively. It is
obvious that the symmetry of the energy levels about $E=0$ is broken as an
effect of such rainbow functions structure. In Fig.2(a), we observe the
asymptotic tendency of the energy levels as $\beta \rightarrow 0$ (i.e., the
cosmic string spacetime limit) and as $\beta >>1$. It is obvious that $\beta
\rightarrow 0\Rightarrow E_{n_{r},\ell }=\pm \sqrt{K_{n_{r},\ell }+m^{2}}$,
whereas $\beta >>1\Rightarrow E_{n_{r},\ell }\rightarrow 0$ in the upper
half and $E_{n_{r},\ell }\sim -2\beta K_{n_{r},\ell }$ in the lower half of
the energy levels. Moreover, Figure 2(b) shows similar trend of asymptotic
convergence as an effect of the magnetic field for a given value of the
rainbow gravity parameter $\beta =0.7$.
\begin{figure}[!ht]  
\centering
\includegraphics[width=0.35\textwidth]{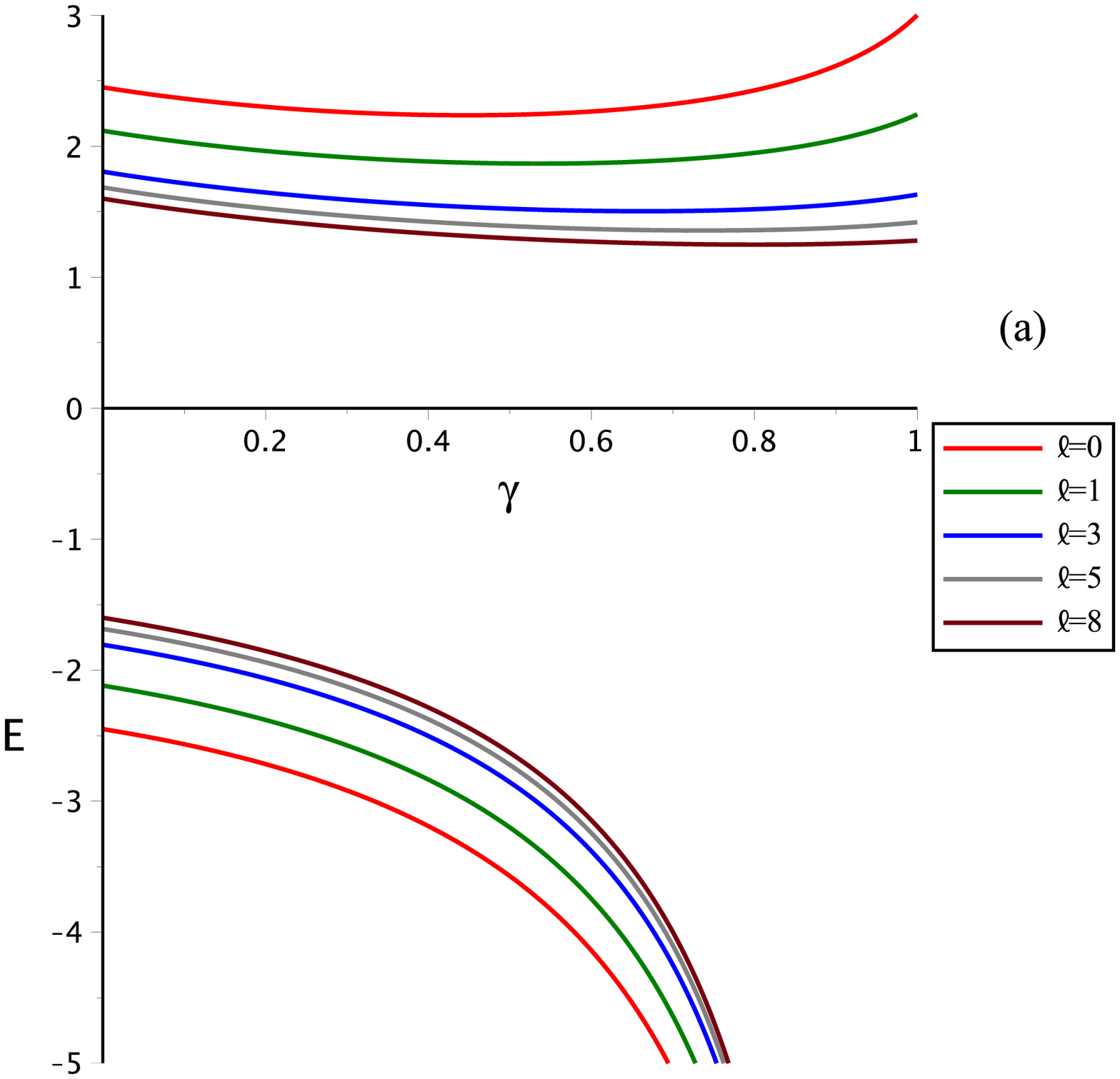}
\includegraphics[width=0.35\textwidth]{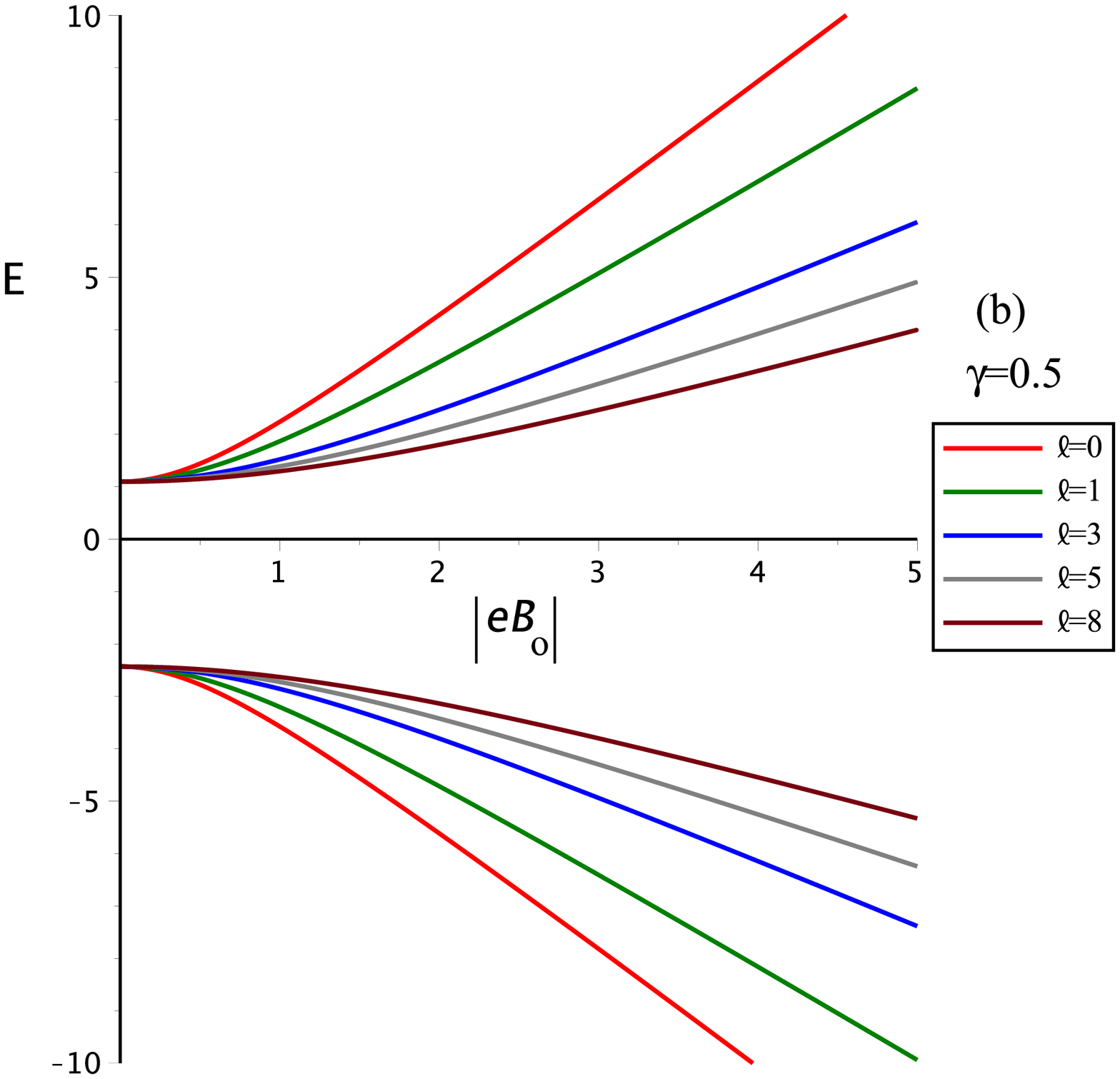} 
\caption{\small 
{ The energy levels of (\ref{e26}), using $\alpha =1/4$, $%
m=k_{z}=1$, so that (a) shows $E$ against $\gamma =\epsilon m/E_{p}<1$ for $%
|eB_{\circ }|=1$, $n_{r}=2$, $\ell =0,1,3,5,8$, and (b) shows $E$ against $%
|eB_{\circ }|$ for $\gamma =0.5$, $n_{r}=2$, $\ell =0,1,3,5,8$.}}
\label{fig3}
\end{figure}%
Yet, in this case, it is more rapid
since $K_{n_{r},\ell }\sim \tilde{B}^{2}$ for $\left\vert eB_{\circ
}\right\vert >>1$ and consequently all energy levels cluster around $%
E_{n_{r},\ell }\sim 1/2\beta $ ($=5$ for $\beta =0.1$ used in the figure) in
the upper half and $E_{n_{r},\ell }\sim -2a\tilde{B}^{2}$ (where $0<a=\frac{1%
}{4}\left[ 1-\frac{\tilde{\ell}^{2}}{\left( n_{r}+|\tilde{\ell}|+\frac{1}{2}%
\right) ^{2}}\right] <\frac{1}{4}$ \ ) in the lower half. This effect is
obvious from the energy levels in Eq.(\ref{e25}), as the first negative term
increases the negativity of the energy levels and breaks the symmetry of the
energy part of the second term.%

\subsection{The set of rainbow functions $g_{_{0}}\left( y\right)
=g_{_{1}}\left( y\right) =1/\left( 1-\epsilon y\right) $}

Upon the substitution of $g_{_{0}}\left( y\right) =g_{_{1}}\left( y\right)
=1/\left( 1-\epsilon y\right) $ in Eq.(\ref{e21}) we obtain%
\begin{equation}
E^{2}-K_{n_{r},\ell }=\left( 1-\epsilon \frac{E}{E_{p}}\right)
^{2}m^{2}\Longrightarrow E=\frac{-m\gamma \pm \sqrt{K_{n_{r},\ell }\left(
1-\gamma ^{2}\right) +m^{2}}}{1-\gamma ^{2}};\;\gamma =\frac{\epsilon m}{%
E_{p}}<1.  \label{e26}
\end{equation}%
In Figures 3(a) we plot the energy levels against $\gamma =\epsilon
m/E_{p}<1 $ to observe the rainbow gravity effect. We clearly see that the
symmetry in the energy levels is broken as an effect of the first term $%
\left[ -m\gamma /\left( 1-\gamma ^{2}\right) \right] $ in Eq.(\ref{e26}). In
Figure 3(b) the energy levels are plotted against $\left\vert eB_{\circ
}\right\vert $ so that the magnetic field effect on the energy levels is
shown.%

\subsection{The set of rainbow functions $g_{_{0}}( y) =( e^{%
\epsilon y}-1) /\epsilon y$ and $g_{_{1}}\left(
y\right) =1$}

We now use $g_{_{0}}( y) =( e^{\epsilon y}-1)/\epsilon y$ and $g_{_{1}}\left( y\right) =1$ so that Eq.(\ref{e21}) implies%
\begin{equation}
E^{2}\left( \frac{e^{\epsilon E/E_{p}}-1}{\epsilon E/E_{p}}\right)
^{2}-m^{2}=K_{n_{r},\ell }\Longrightarrow E=\frac{1}{2\beta }\ln \left( 1\pm 
\sqrt{4\beta ^{2}\left( K_{n_{r},\ell }+m^{2}\right) }\right) ;\;\beta =\frac{\epsilon }{2E_{p}}  \label{e27}
\end{equation}%
One should notice that $\left( \frac{e^{\epsilon E/E_{p}}-1}{\epsilon E/E_{p}}\right)^2 \rightarrow 1$ as $\epsilon\rightarrow 0$ (i.e., at the cosmic string spacetime limit) and the energy levels retrieve their symmetry about the $E=0$ line. In Figure 4(a) we plot the energy levels against $\beta =\epsilon /2E_{p}$ and observe eminent clustering in the positive energies as $\beta $ grows up
from just above zero (i.e., $\beta \geq 0.001$), whereas the negative
energies are rapidly pushed further into the negative energy region. In
Figures 4(b) we show the effect of the magnetic field on the energy levels.%
\begin{figure}[!ht]  
\centering
\includegraphics[width=0.35\textwidth]{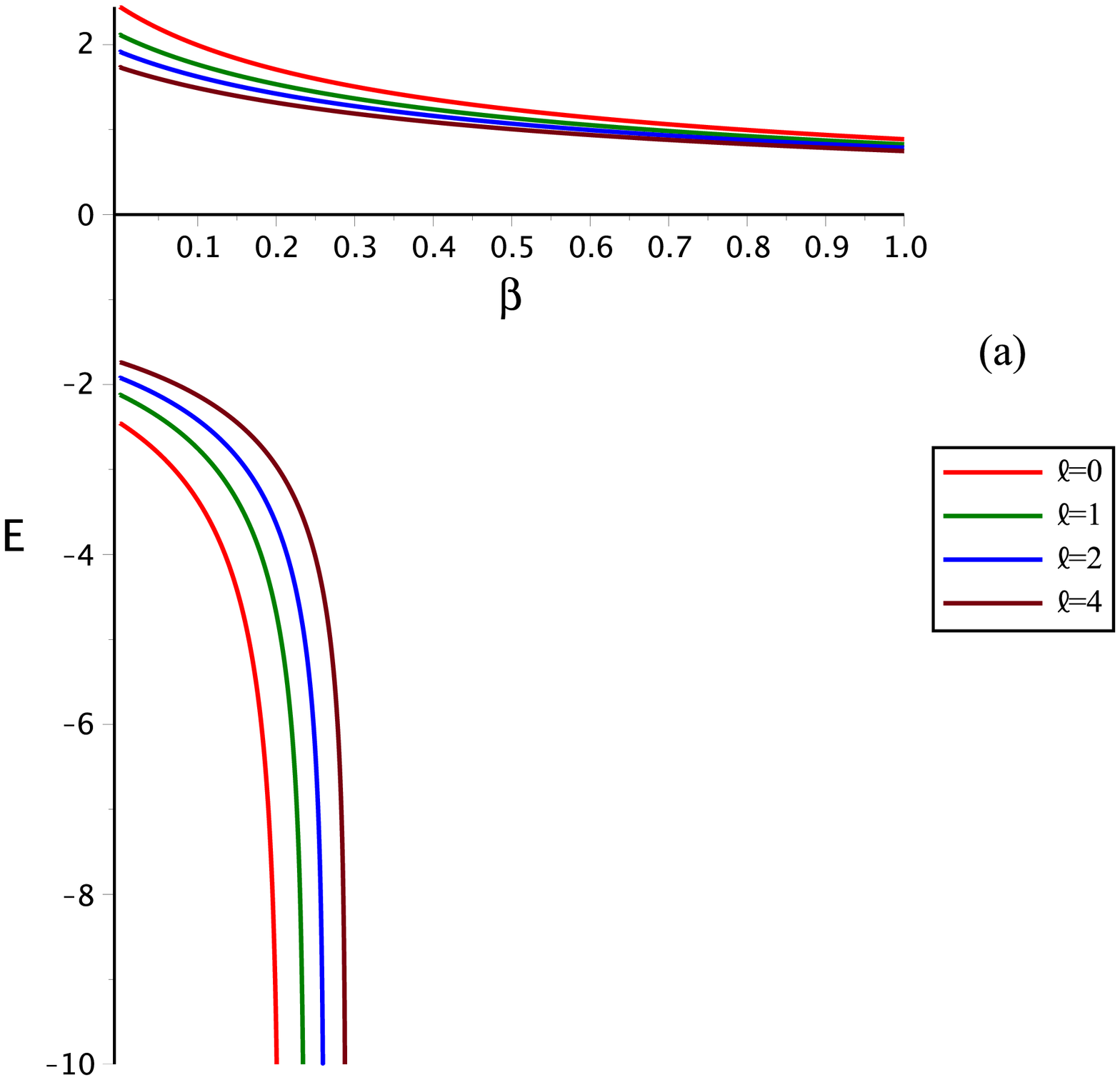}
\includegraphics[width=0.35\textwidth]{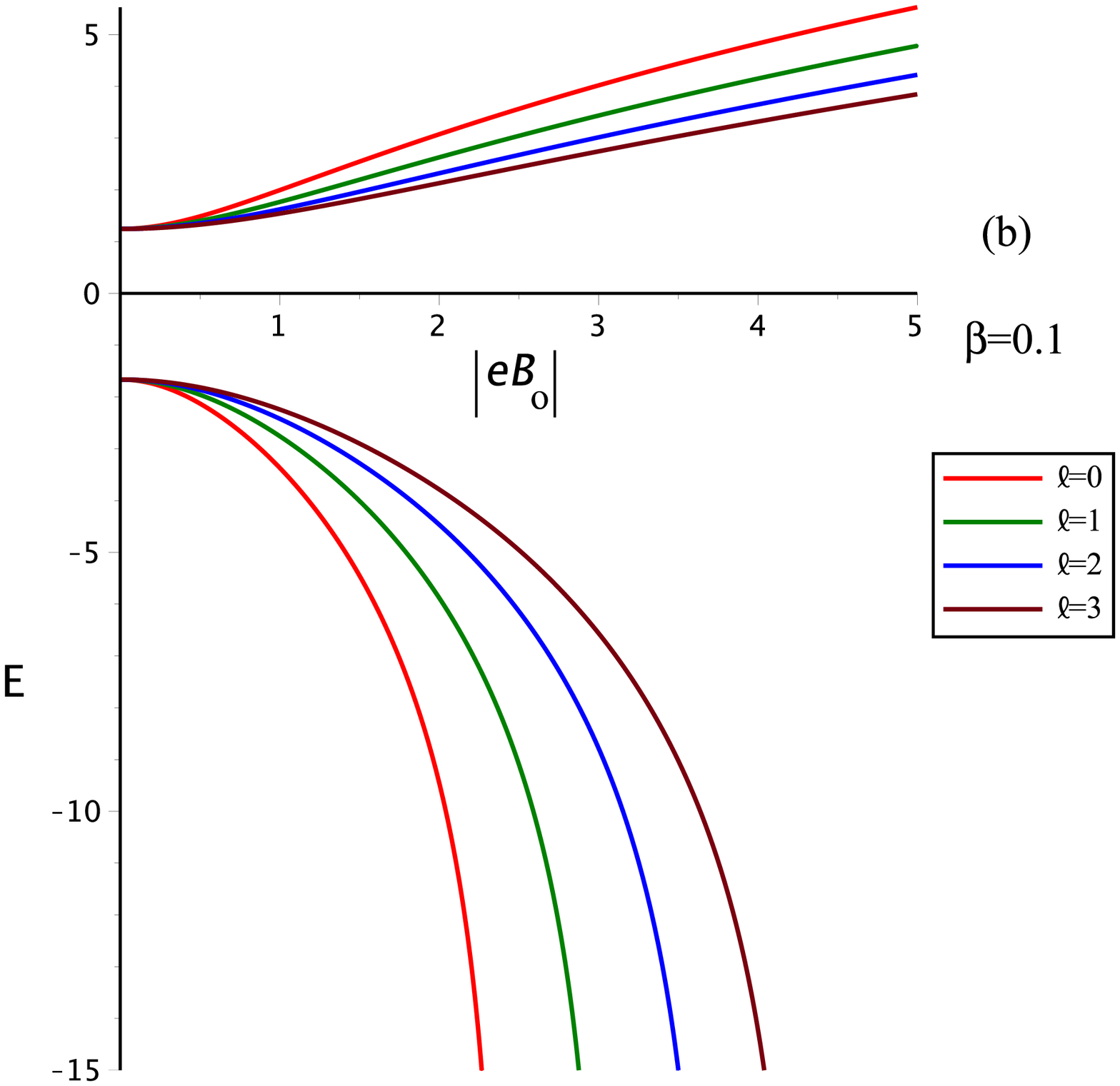} 
\caption{\small 
{ The energy levels of (\ref{e27}), using $\alpha =1/4$, $%
m=k_{z}=1$, so that (a) shows $E$ against $\beta =\epsilon /2E_{p}$ for $%
|eB_{\circ }|=1$, $n_{r}=2$, $\ell =0,1,2,4$, and (b) shows $E$ against $%
|eB_{\circ }|$ for $\beta =0.1$, $n_{r}=2$, $\ell =0,1,2,3$.}}
\label{fig4}
\end{figure}%

\section{PDM KG-particles in cosmic string rainbow gravity spacetime and a uniform magnetic field}
In this section, we consider a positive-valued dimensionless scalar
multiplier in the form of $f\left( r\right) =\exp \left( 4\eta r\right) $
in Eq.(\ref{e10}) so that $M( r) =\eta ^{2}+\eta /r$. This
would, in turn, imply that Eq.(\ref{e16}) reads%
\begin{equation}
\left\{ \partial _{r}^{2}-\frac{\left( \tilde{\ell}^{2}-1/4\right) }{r^{2}}+%
\frac{\left( -\eta +\tilde{\ell}\,\tilde{B}\right) }{r}+\tilde{\lambda}%
\right\} R\left( r\right) =0,\;  \label{e28}
\end{equation}%
where%
\begin{equation}
\tilde{\lambda}=\frac{g_{_{0}}\left( y\right) ^{2}E^{2}-m^{2}}{%
g_{_{1}}\left( y\right) ^{2}}-\left( k_{z}^{2}+\frac{\tilde{B}^{2}}{4}%
\right) -\eta ^{2},  \label{e29}
\end{equation}%
In this case, its exact solution is in the form of%
\begin{equation}
R( r) =C\,r^{|\tilde{\ell}|+1/2}\exp ( -i\sqrt{\tilde{\lambda%
}}r) \,_{1}F_{1}\left( \frac{1}{2}+|\tilde{\ell}|-\frac{|-\eta +\tilde{%
\ell}\,\tilde{B}|}{2i\sqrt{\tilde{\lambda}}},1+2|\tilde{\ell}|,2i\sqrt{%
\tilde{\lambda}}r\right) ,  \label{e30}
\end{equation}%
which takes the form of a polynomial of degree $n_{r}\geq 0$ for 
\begin{equation}
\frac{1}{2}+|\tilde{\ell}|-\frac{|-\eta +\tilde{\ell}\,\tilde{B}|}{2i\sqrt{%
\tilde{\lambda}}}=-n_{r}\Rightarrow \frac{1}{2}+|\tilde{\ell}|-\frac{|-\eta +%
\tilde{\ell}\,\tilde{B}|}{2i\sqrt{\tilde{\lambda}}}=-n_{r}\Rightarrow 2i%
\sqrt{\tilde{\lambda}}=\frac{|-\eta +\tilde{\ell}\,\tilde{B}|}{\tilde{n}};\;%
\tilde{n}=n_{r}+|\tilde{\ell}|+\frac{1}{2}.  \label{e301}
\end{equation}%
Consequently, the eigen energies and wavefunctions, with $\tilde{\eta}=-\eta +%
\tilde{\ell}\,\tilde{B}$, are given, respectively, by%
\begin{equation}
\tilde{\lambda}_{n_{r},\ell }=-\frac{\tilde{\eta}^{2}}{4\tilde{n}^{2}},
\label{e31}
\end{equation}
and 
\begin{equation}
\psi \left( r\right) =\frac{R\left( r\right) }{\sqrt{r}}=C\,r^{|\tilde{\ell}%
|}\exp \left( -\frac{|\tilde{\eta}|}{2\tilde{n}}r\right) \,_{1}F_{1}\left(
-n_{r},1+2|\tilde{\ell}|,\frac{|\tilde{\eta}|}{\tilde{n}}r\right) .
\label{e32}
\end{equation}%
Hence, Eq.s (\ref{e29}) and (\ref{e31}) along with (\ref{e15}) would result 
\begin{equation}
g_{_{0}}\left( y\right) ^{2}E^{2}-m^{2}=g_{_{1}}\left( y\right) ^{2}\,\tilde{%
K}_{n_{r},\ell };\;\tilde{K}_{n_{r},\ell }=\left[ k_{z}^{2}+\frac{\tilde{B}%
^{2}}{4}\left( 1-\frac{\tilde{\ell}^{2}}{\tilde{n}^{2}}\right) +\eta
^{2}\left( 1-\frac{1}{4\tilde{n}^{2}}\right) +\frac{\eta \tilde{\ell}\,%
\tilde{B}}{2\tilde{n}^{2}}\right] .  \label{e33}
\end{equation}%
Evidently, the last term of $\tilde{K}_{n_{r},\ell }$ in (\ref{e33}) would
lift the degeneracies associated with $\ell =\pm |\ell |\neq 0$ and states with
both $\ell $ values would reappear in the spectrum, therefore. One should,
moreover, notice that as $\tilde{B}\rightarrow 0$ our $\tilde{K}_{n_{r},\ell
}\rightarrow k_{z}^{2}+\eta ^{2}\left( 1-1/4\tilde{n}^{2}\right) $ and
consequently states with a specific $\ell =\pm |\ell |$ will emerge from the
same $\tilde{B}=0$ and split as $\tilde{B}$ grows up from zero. We may now
discuss the effects of rainbow gravity for different rainbow functions on
the energy levels under the current metaphoric PDM KG-particles in cosmic
string spacetime and uniform magnetic field of Eq.(\ref{e33}).

\subsection{The set of rainbow functions $g_{_{0}}\left( y\right) =1$, $%
g_{_{1}}\left( y\right) =\sqrt{1-\epsilon y^{n}}$}

For rainbow functions $g_{_{0}}\left( y\right) =1$ and $g_{_{1}}\left(
y\right) =\sqrt{1-\epsilon y^{2}}$ (i.e., $n=2$), Eq.(\ref{e33}) would result%
\begin{equation}
E^{2}-m^{2}=\left( 1-\epsilon \frac{E^{2}}{E_{p}^{2}}\right) \tilde{K}%
_{n_{r},\ell }\Longrightarrow E=\pm \sqrt{\frac{\tilde{K}_{n_{r},\ell }+m^{2}%
}{1+\delta \,\tilde{K}_{n_{r},\ell }}};\;\delta =\frac{\epsilon }{E_{p}^{2}}.
\label{e34}
\end{equation}%

In Figures 5(a), we plot the energies against $\delta =\epsilon /E_{p}^{2}$.
We observe that both halves of the energy levels asymptotically converge to $%
E=0$ value as $\delta >>1$ for a fixed value of the PDM parameter $\eta $.
In Fig. 5(b), we plot the energies against the PDM parameter $\eta $ and
notice that the degeneracies associated with $\ell =\pm |\ell |$ are removed
as $\eta $ increases from zero (i.e., it is obvious that all states with .$%
\ell =\pm |\ell |$ emerge form the same $\eta =0$ point and split as $\eta $
grows up). However, as $\eta >>1$ we observe that the energy levels merge
into $E_{n_{r},\ell }\sim \pm \sqrt{1/\delta }=\pm \sqrt{10}$ for $\delta
=0.1.$In Figure 5(c), moreover, we plot the energies against $|eB_{\circ }|$ and observe that the energy levels with a specific $%
\ell =\pm |\ell |$ split as $|eB_{\circ }|$ increases
from the zero value, and for $|eB_{\circ }| >>1$ the
energy levels cluster and merge into $E_{n_{r},\ell }\sim \pm \sqrt{1/\delta 
}$ (this is observed in Figure 5(c), i.e., as $|eB_{\circ
}| >>1$ the energies $E_{n_{r},\ell }\sim \pm \sqrt{1/\delta }=\pm 
\sqrt{10}$ for $\delta =0.1$). We again observe that such rainbow functions structure the energy levels are destined
to be within the range $\sqrt{( k_{z}^{2}+m^{2}) /( 1+\delta
k_{z}^{2}) }\leq |E_{n_{r},\ell }|\,\leq \sqrt{1/\delta }=E_{p}/\sqrt{%
\epsilon }$.

The second set of the rainbow functions $g_{_{0}}\left( y\right) =1$ and $%
g_{_{1}}\left( y\right) =\sqrt{1-\epsilon y}$ (i.e., $n=1$), on the other
hand, implies (using Eq. (\ref{e33})) that%
\begin{equation}
E^{2}-m^{2}=\left( 1-\epsilon \frac{E}{E_{p}}\right) \tilde{K}_{n_{r},\ell
}\Longrightarrow E=-\beta \,\tilde{K}_{n_{r},\ell }\pm \sqrt{\beta ^{2}%
\tilde{K}_{n_{r},\ell }^{2}+\tilde{K}_{n_{r},\ell }+m^{2}};\;\beta =\frac{%
\epsilon }{2E_{p}}.  \label{e35}
\end{equation}%

In Figures 6(a) and (b), we plot the energy levels against $\beta =\epsilon
/2E_{p}$ and $|eB_{\circ }|$, respectively. It is
obvious that the symmetry of the energy levels about $E=0$ is broken as an
effect of such rainbow functions structure. In Fig.6(a) we notice that the
clustering around the $S$-state with $\ell =0$ only occurs for the upper
half of the KG-energies (i.e., positive energies), whereas in the lower half
we observe that the splitting in the energy levels increases as $\beta $
increases from zero. In the lower half of Fig.6(b) we see that the energy
levels separation increases as the magnetic field strength increases from
zero. This effect is obvious from the form of the energy levels in Eq.(\ref%
{e35}), as the first negative term increases the negativity of the energy
levels and breaks the symmetry of the energy part of the second term. In
Fig.6(c) we show the effect of the PDM settings on the energy levels where
the symmetry of the energy levels about $E=0$ is broken because of the
rainbow functions structure.
\begin{figure}[!ht]  
\centering
\includegraphics[width=0.3\textwidth]{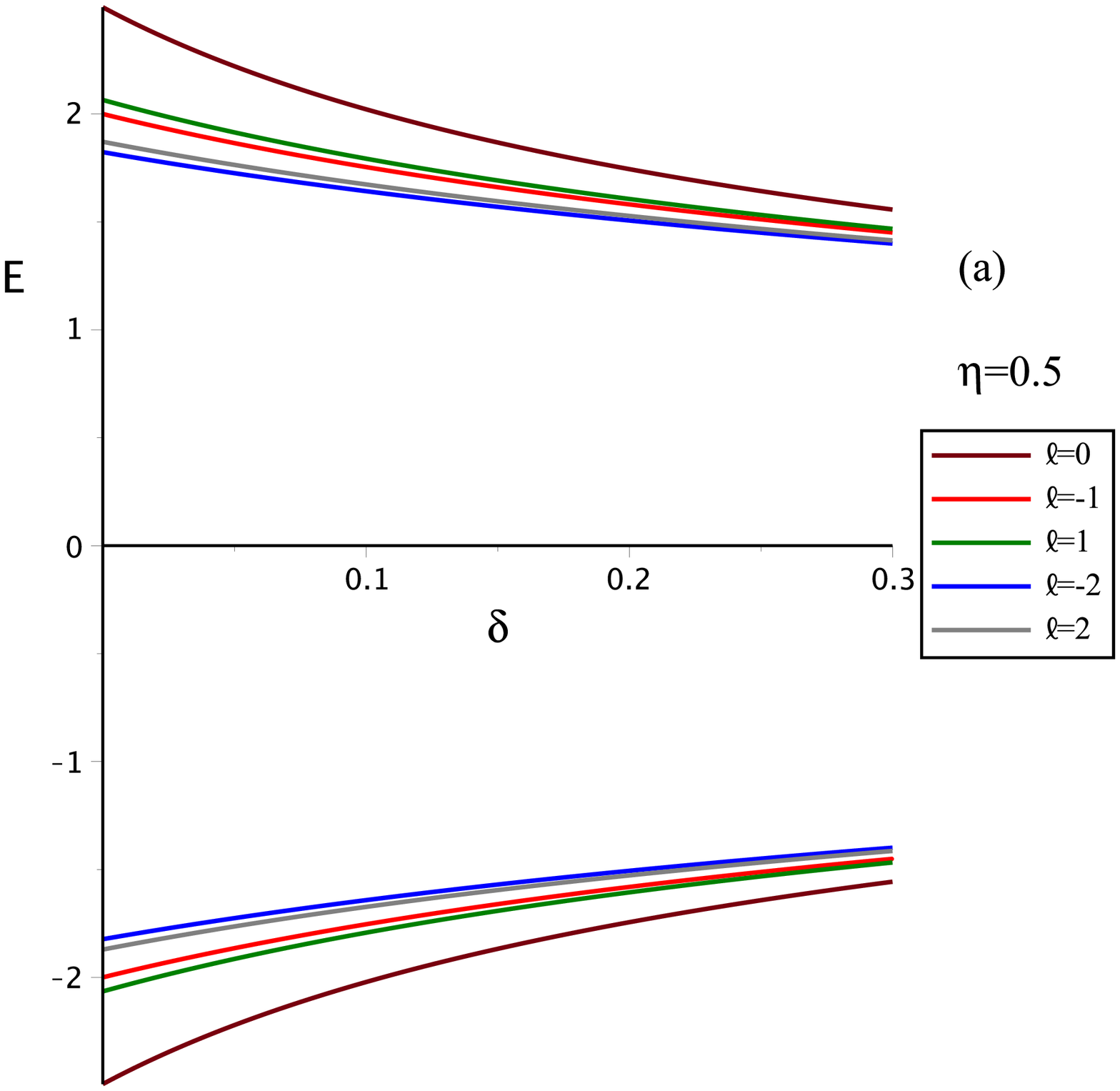}
\includegraphics[width=0.3\textwidth]{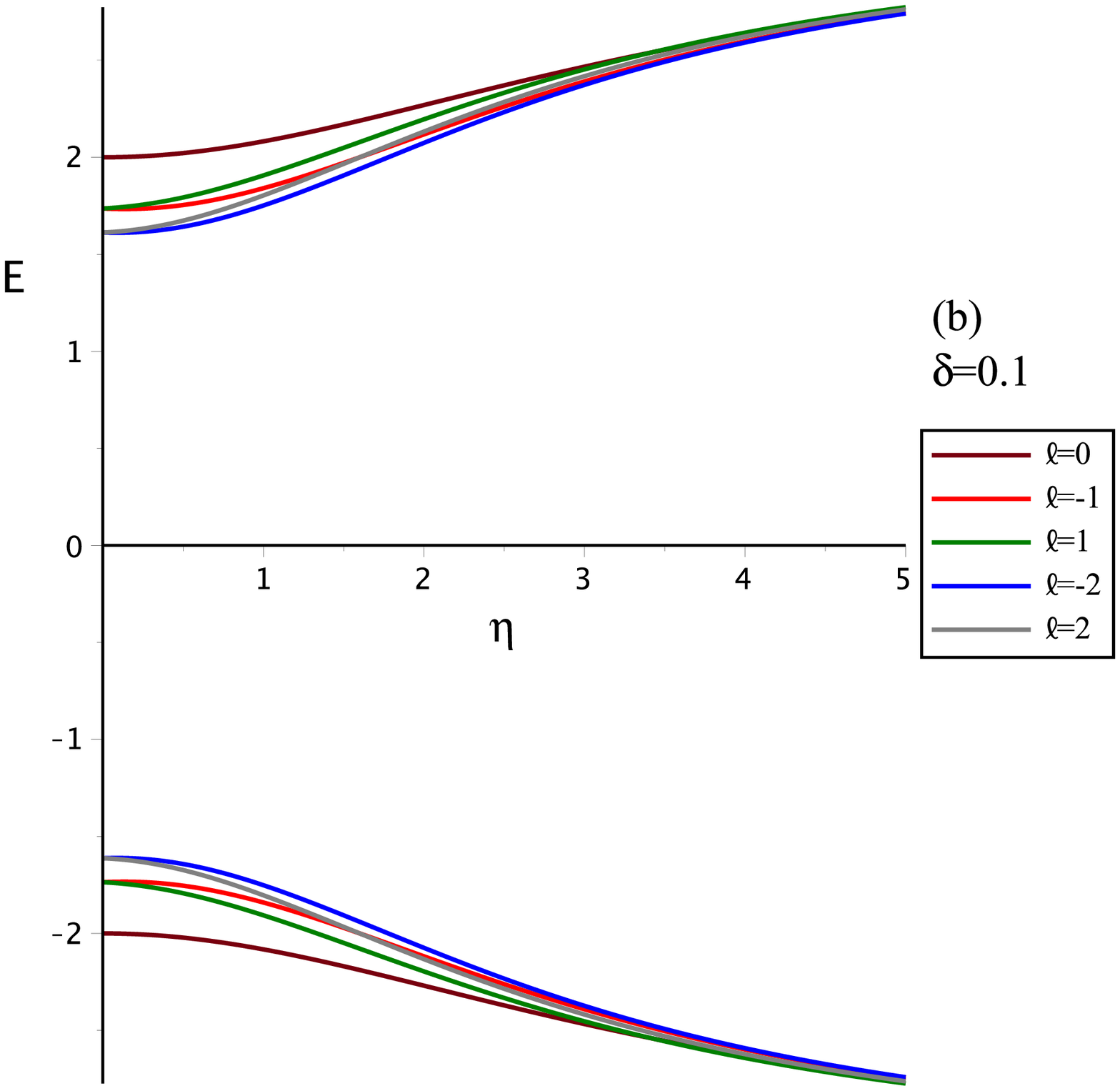}
\includegraphics[width=0.3\textwidth]{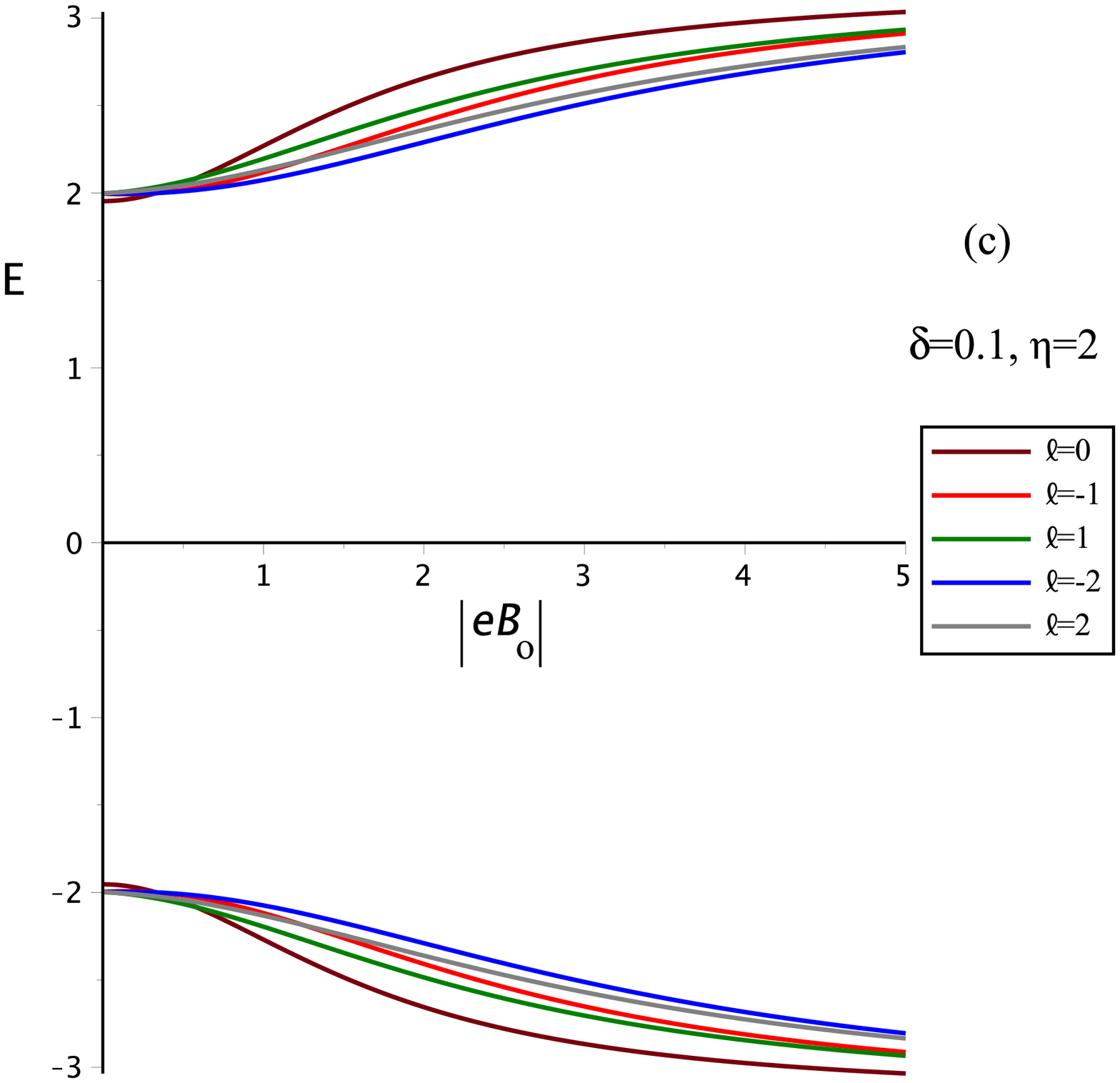} 
\caption{\small 
{ The energy levels of (\ref{e34}), using $m=k_{z}=1$, so that
(a) shows $E$ against $\delta =\epsilon /E_{p}^{2}$ for $\alpha =1/4$, $%
|eB_{\circ }|=1$, $\eta =0.5$, $n_{r}=1$, $\ell =0,\pm 1,\pm 2$, (b) shows $%
E $ against $\eta $ for $\alpha =1/4$, $|eB_{\circ }|=1$, $\delta =0.1$, $%
n_{r}=1$, $\ell =0,\pm 1,\pm 2$, and (c) shows $E$ against $|eB_{\circ }|$
for $\alpha =1/4$, $\eta =2$, $\delta =0.1$, $n_{r}=1$, $\ell =0,\pm 1,\pm 2$.}}
\label{fig5}
\end{figure}%
\begin{figure}[!ht]  
\centering
\includegraphics[width=0.3\textwidth]{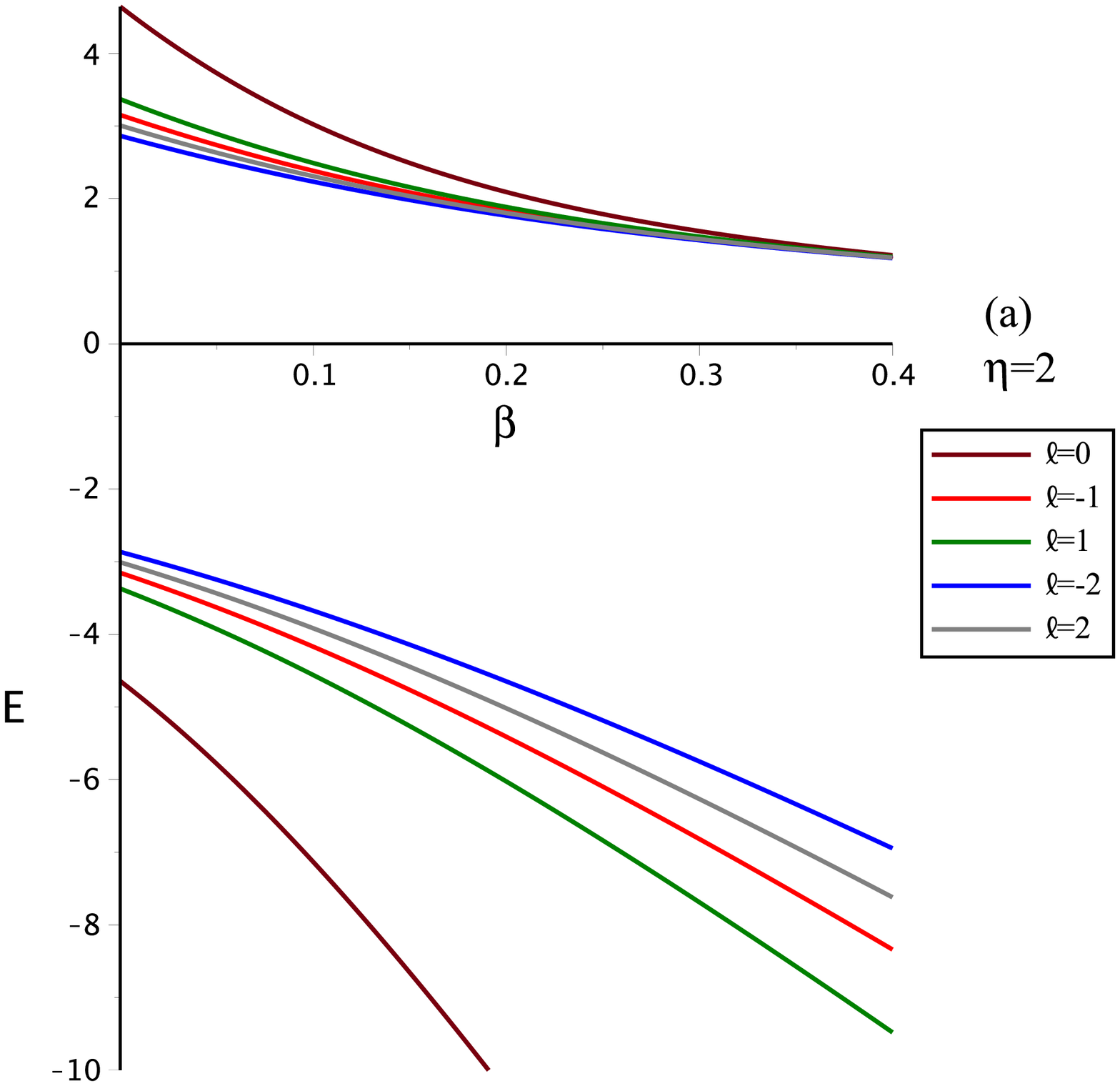}
\includegraphics[width=0.3\textwidth]{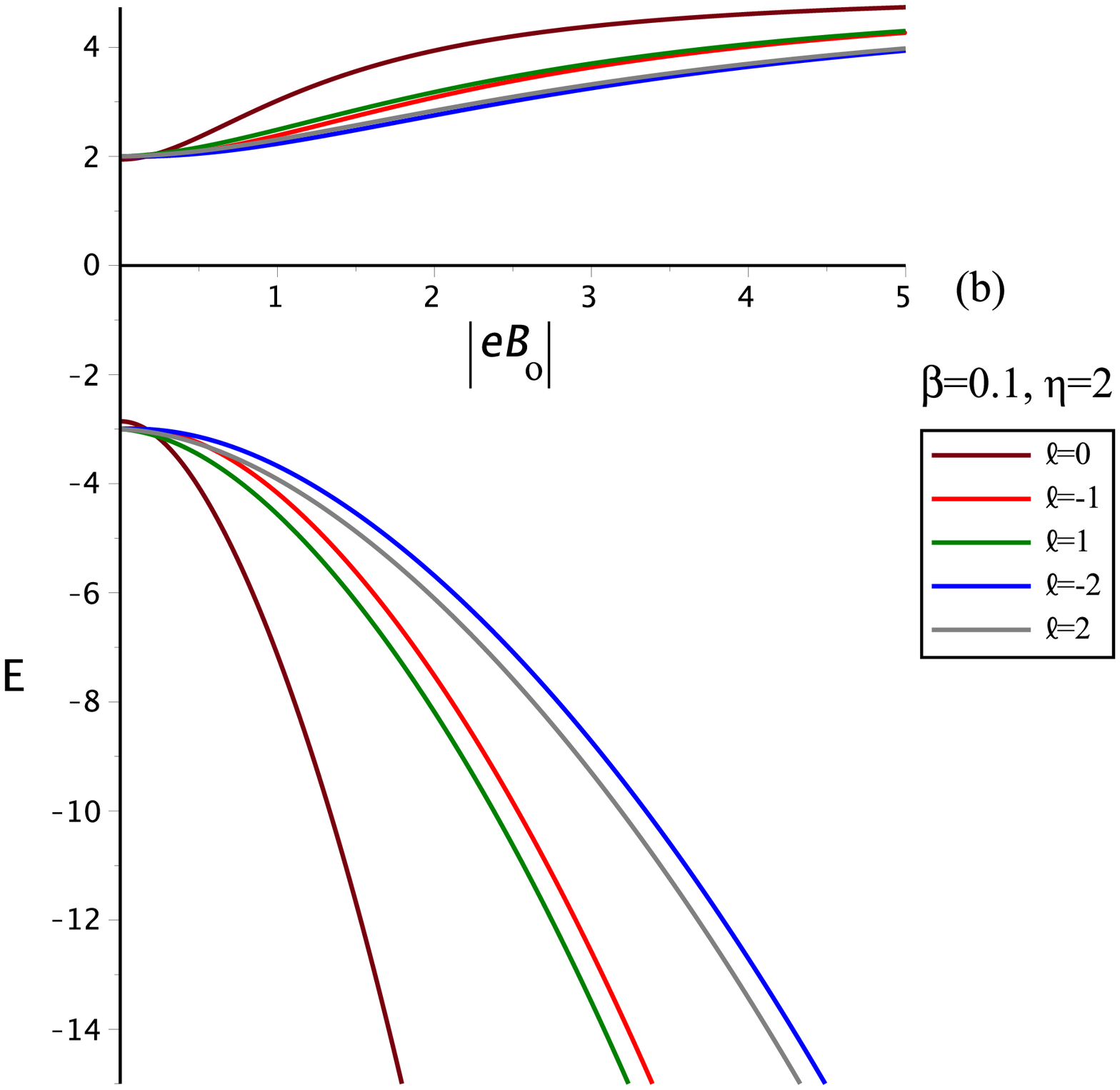}
\includegraphics[width=0.3\textwidth]{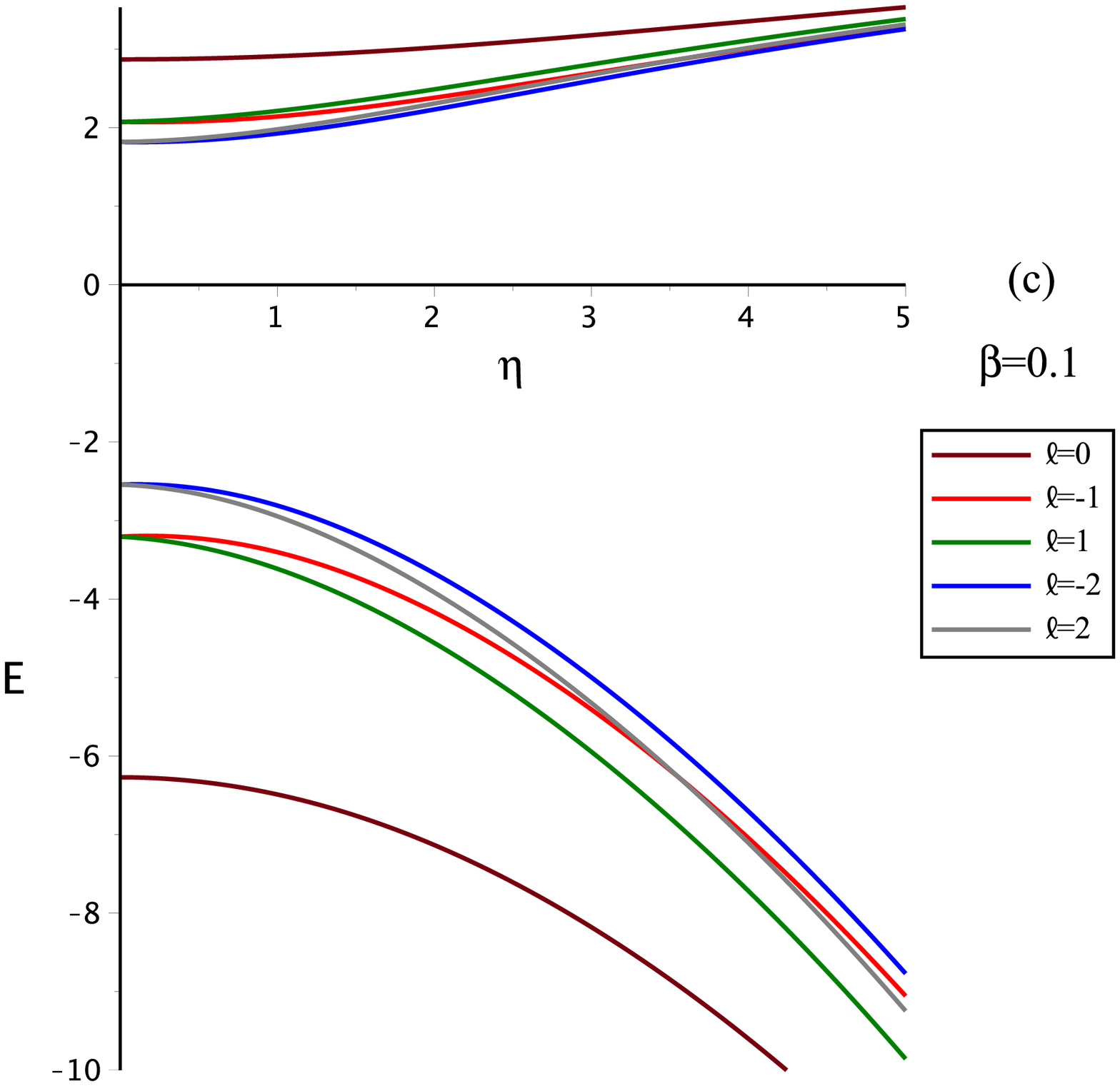} 
\caption{\small 
{ The energy levels of (\ref{e35}), using $\alpha =1/8$, $%
m=k_{z}=1$, so that (a) shows $E$ against $\beta =\epsilon /2E_{p}$ for $%
|eB_{\circ }|=1$, $\eta =2$, $n_{r}=1$, $\ell =0,\pm 1,\pm 2$, (b) shows $E$
against $|eB_{\circ }|$ for $\eta =2$, $\beta =0.1$, $n_{r}=1$, $\ell =0,\pm
1,\pm 2$, and (c) shows $E$ against $\eta $ for $|eB_{\circ }|=1$, $\beta
=0.1$, $n_{r}=1$, $\ell =0,\pm 1,\pm 2$.}}
\label{fig6}
\end{figure}%

\subsection{The set of rainbow functions $g_{_{0}}\left( y\right)
=g_{_{1}}\left( y\right) =1/\left( 1-\epsilon y\right) $}

For the rainbow functions $g_{_{0}}\left( y\right) =g_{_{1}}\left( y\right)
=1/\left( 1-\epsilon y\right) $, Eq.(\ref{e33}) yields%
\begin{equation}
E^{2}-\tilde{K}_{n_{r},\ell }=\left( 1-\epsilon \frac{E}{E_{p}}\right)
^{2}m^{2}\Longrightarrow E=\frac{-m\gamma \pm \sqrt{\tilde{K}_{n_{r},\ell
}\left( 1-\gamma ^{2}\right) +m^{2}}}{1-\gamma ^{2}};\;\gamma =\frac{%
\epsilon m}{E_{p}}<1.  \label{e36}
\end{equation}%
In Figures 7(a) we plot the energy levels against $\gamma =\epsilon m/E_{p}$
to observe the rainbow gravity effect. We clearly see that the symmetry in
the energy levels is broken as an effect of the first term $\left[ -m\gamma
/\left( 1-\gamma ^{2}\right) \right] $ in Eq.(\ref{e36}). In Figures 7(b)
the energy levels are plotted against $|eB_{\circ }|$ so
that the magnetic field effect on the energy levels is shown, and in 7(c) we
show the effect of the PDM settings on the energy levels.
\begin{figure}[!ht]  
\centering
\includegraphics[width=0.3\textwidth]{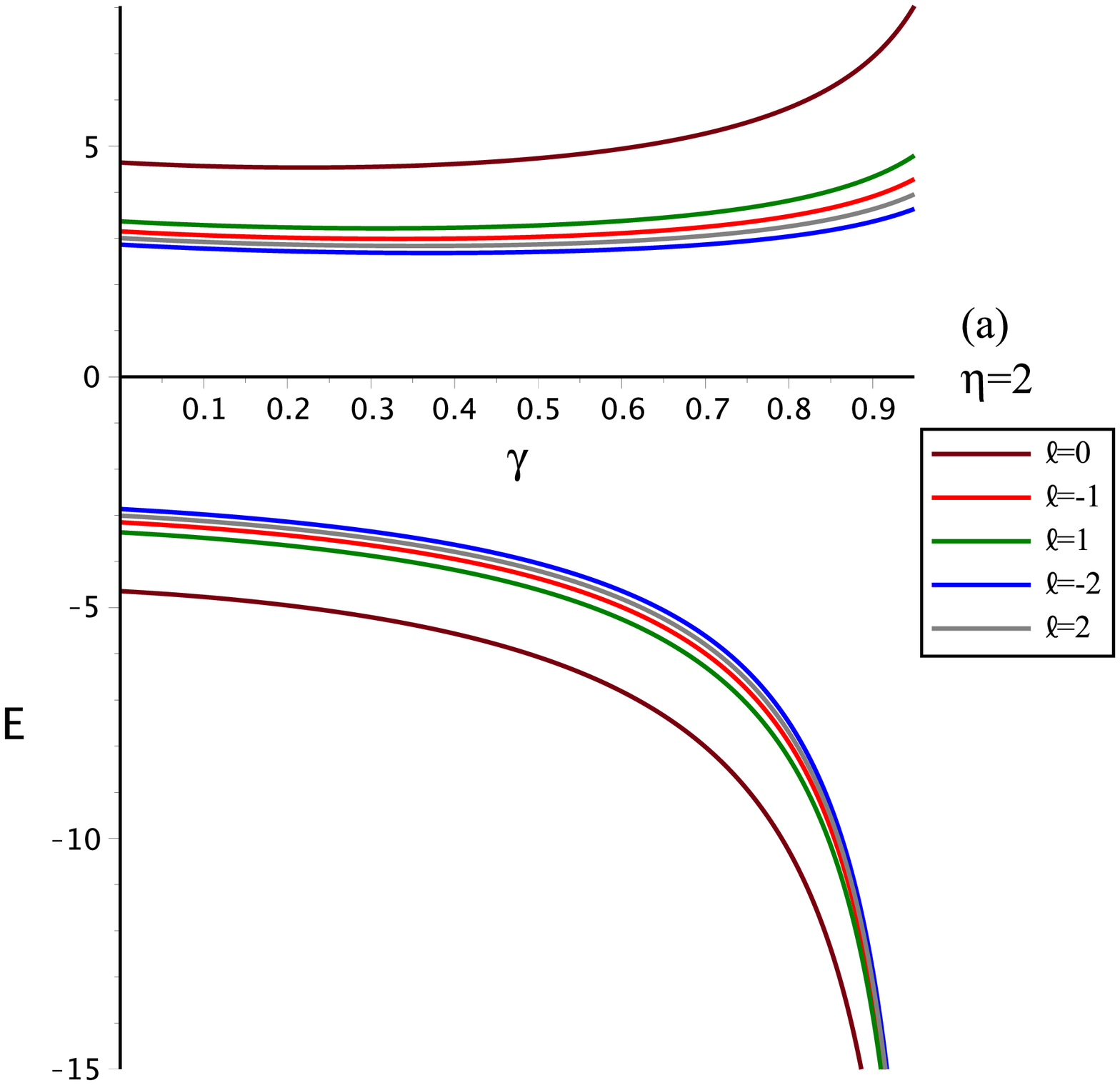}
\includegraphics[width=0.3\textwidth]{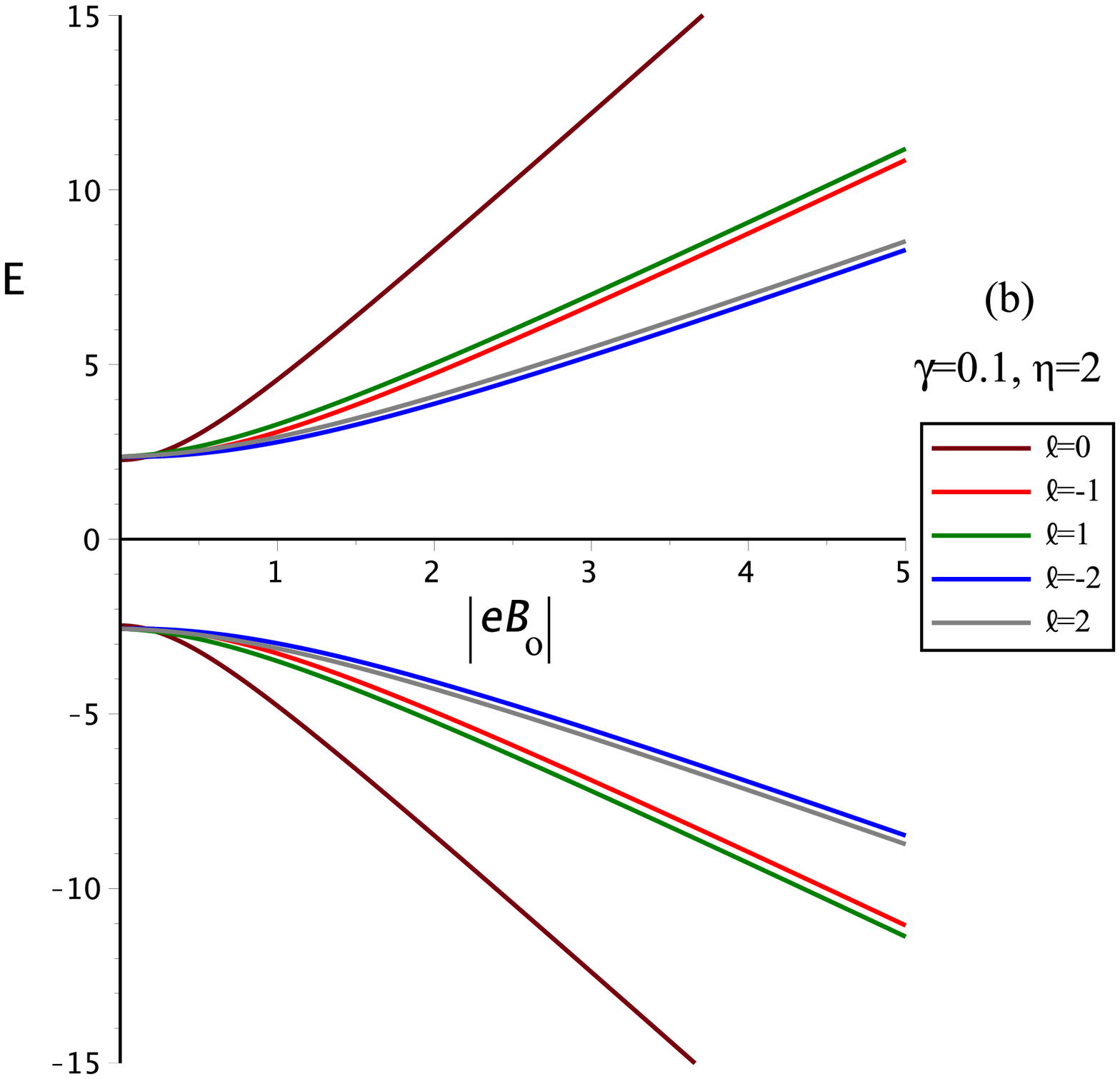}
\includegraphics[width=0.3\textwidth]{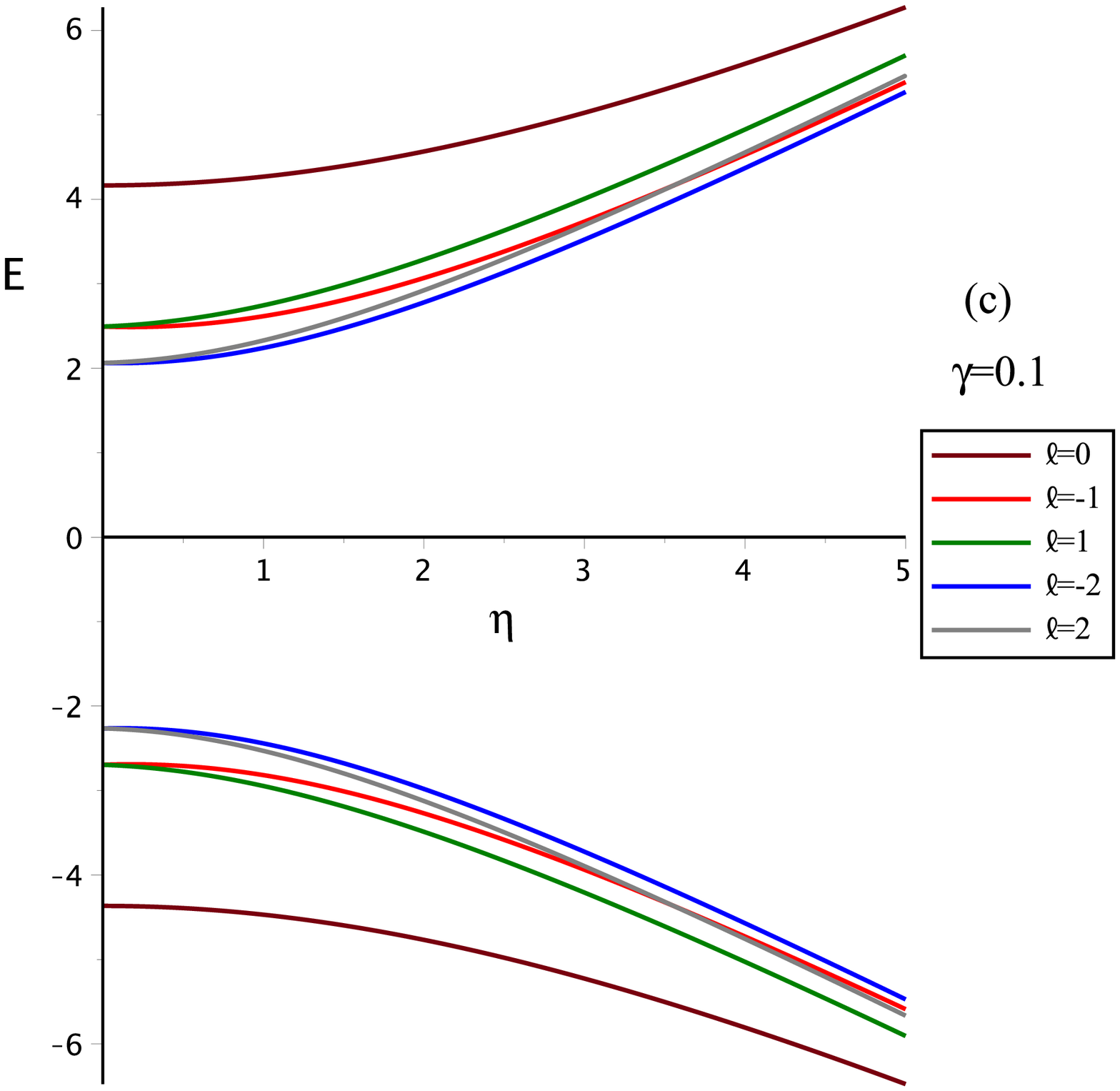} 
\caption{\small 
{ The energy levels of (\ref{e36}), using $\alpha =1/8$, $%
m=k_{z}=1$, so that (a) shows $E$ against $\gamma =\epsilon m/E_{p}<1$ for $%
|eB_{\circ }|=1$, $\eta =2$, $n_{r}=1$, $\ell =0,\pm 1,\pm 2$, (b) shows $E$
against $|eB_{\circ }|$ for $\eta =2$, $\gamma =0.1$, $n_{r}=1$, $\ell
=0,\pm 1,\pm 2$, and (c) shows $E$ against $\eta $ for $|eB_{\circ }|=1$, $%
\gamma =0.1$, $n_{r}=1$, $\ell =0,\pm 1,\pm 2$.}}
\label{fig7}
\end{figure}%
\begin{figure}[!ht]  
\centering
\includegraphics[width=0.3\textwidth]{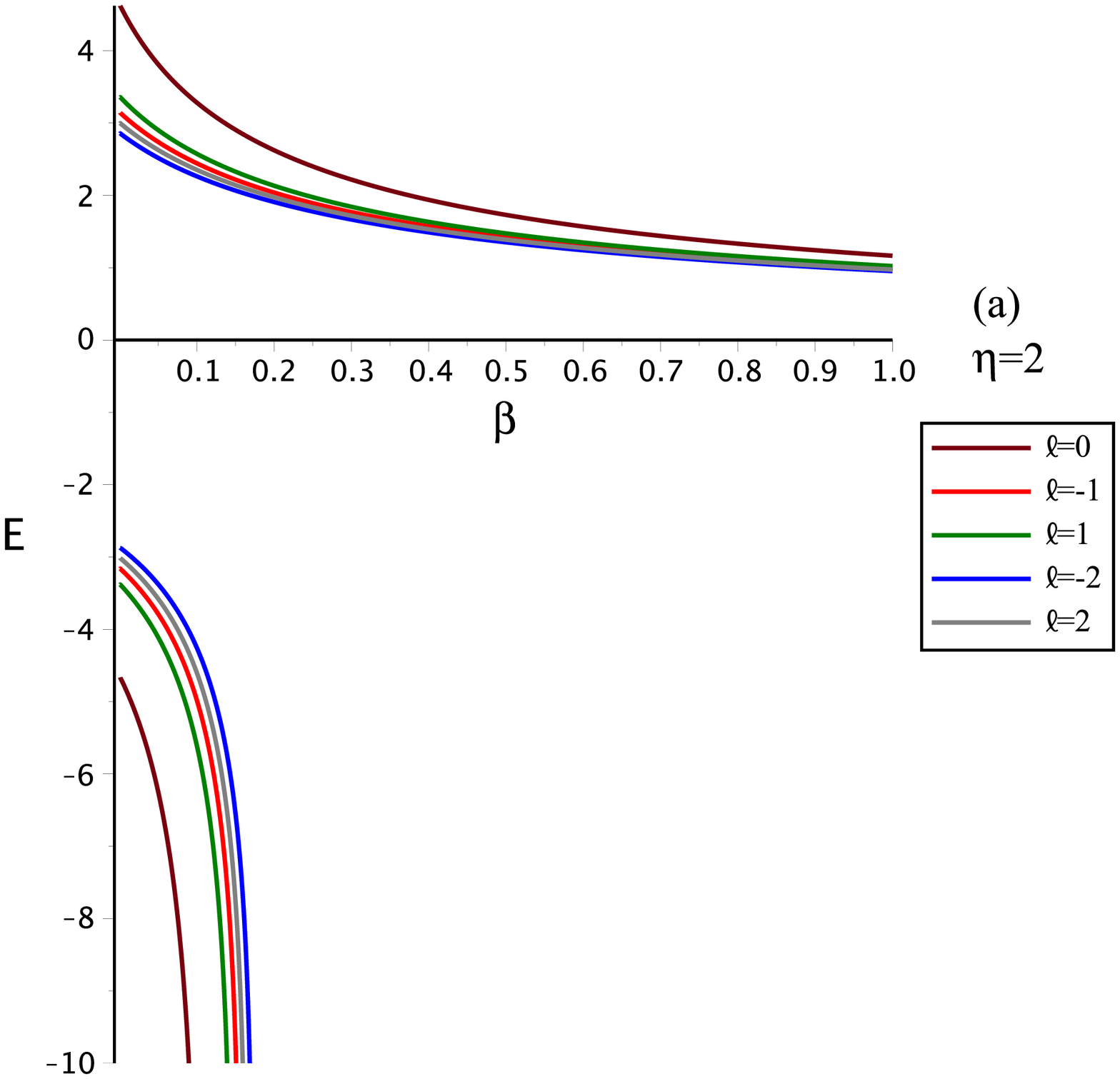}
\includegraphics[width=0.3\textwidth]{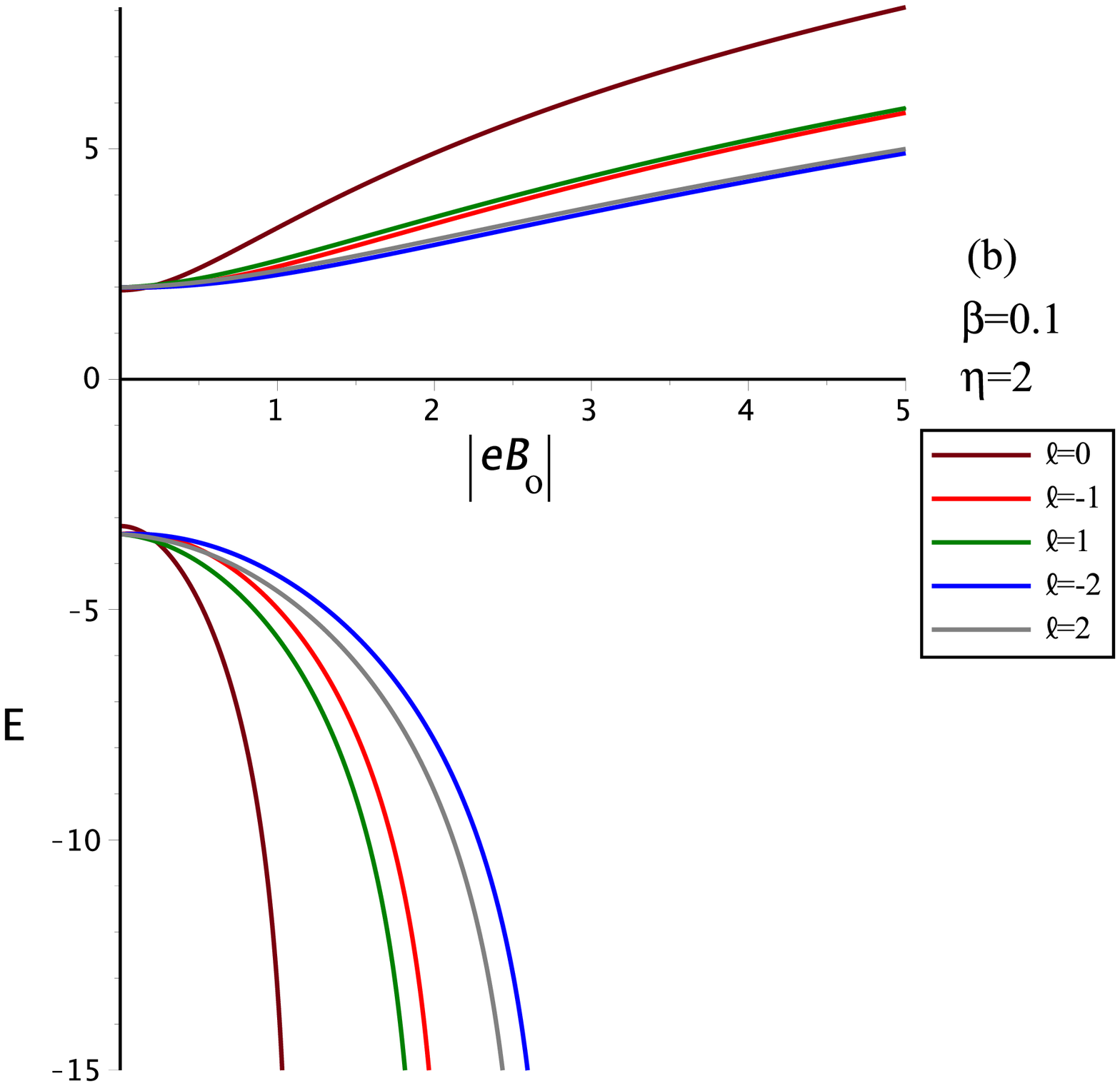}
\includegraphics[width=0.3\textwidth]{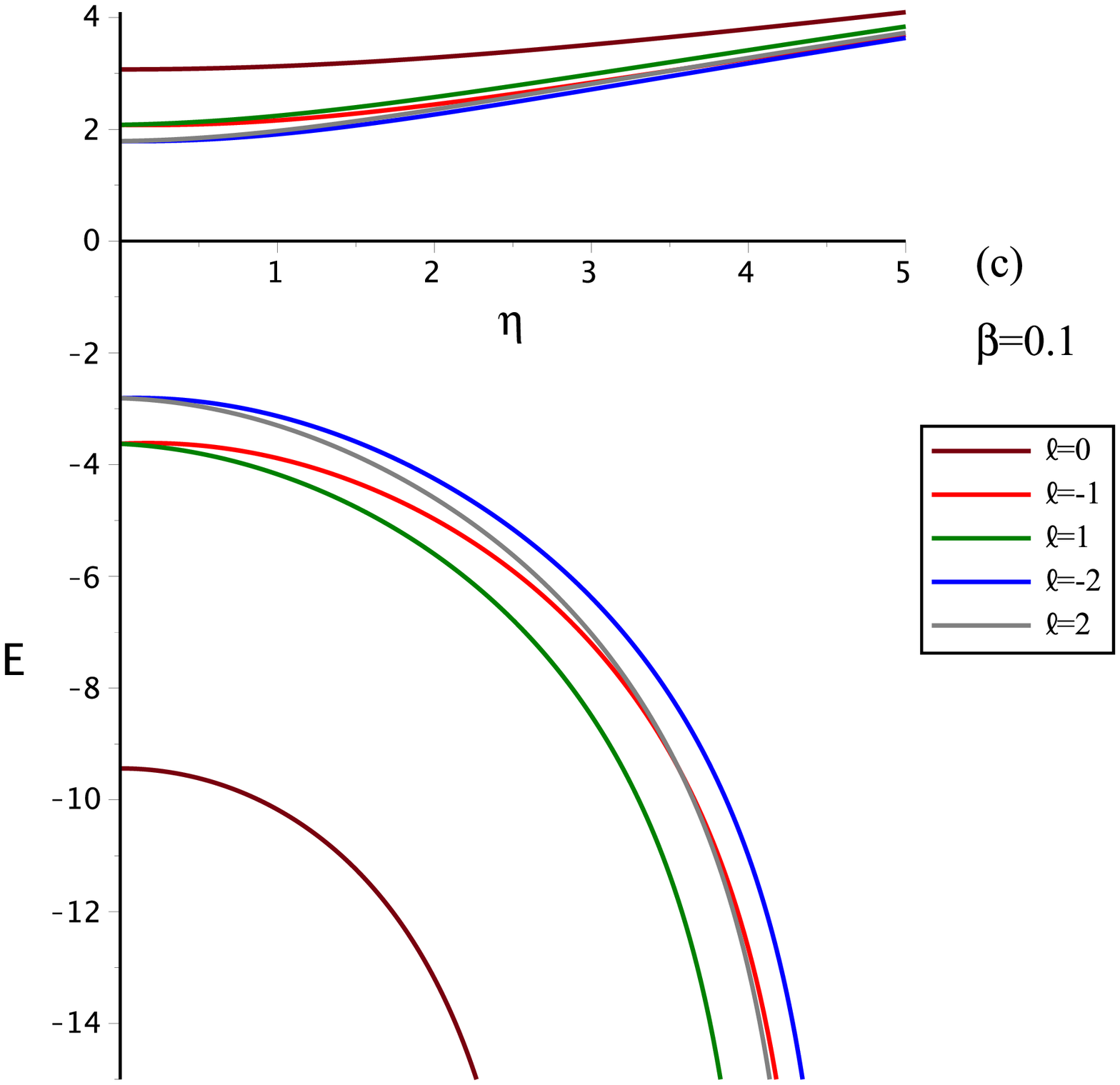} 
\caption{\small 
{ The energy levels of (\ref{e37}), using $\alpha =1/8$, $%
m=k_{z}=1$, so that (a) shows $E$ against $\beta =\epsilon /2E_{p}$ for $%
|eB_{\circ }|=1$, $\eta =2$, $n_{r}=1$, $\ell =0,\pm 1,\pm 2$, (b) shows $E$
against $|eB_{\circ }|$ for $\eta =2$, $\beta =0.1$, $n_{r}=1$, $\ell =0,\pm
1,\pm 2$, and (c) shows $E$ against $\eta $ for $|eB_{\circ }|=1$, $\beta
=0.1$, $n_{r}=1$, $\ell =0,\pm 1,\pm 2$.}}
\label{fig8}
\end{figure}%

\subsection{The set of rainbow functions $g_{_{0}}( y) =( e^{\epsilon y}-1) /\epsilon y$ and $g_{_{1}}\left(
y\right) =1$}

Such rainbow functions structure in Eq.(\ref{e33}) would yield%
\begin{equation}
E^{2}\left( \frac{e^{\epsilon E/E_{p}}-1}{\epsilon E/E_{p}}\right)
^{2}-m^{2}=\tilde{K}_{n_{r},\ell }\Longrightarrow E=\frac{1}{2\beta }\ln
\left( 1\pm \sqrt{4\beta ^{2}\left( \tilde{K}_{n_{r},\ell }+m^{2}\right) }%
\right) ;\;\beta =\frac{\epsilon }{2E_{p}}  \label{e37}
\end{equation}%
In Figure 8(a) we plot the energy levels against $\beta =\epsilon /2E_{p}$
and observe eminent clustering in the positive energies as $\beta $ grows up
from just above zero (i.e., $\beta \geq 0.001$), whereas the negative
energies are rapidly pushed further into the negative energy region. In
Figures 8(b) we show the effect of the magnetic field and in Fig. 8(c) we
show the effect of the PDM settings on the energy levels.
\section{Concluding remarks}
We have considered KG-particles in the cosmic string rainbow gravity spacetime (\ref{e4}) and uniform magnetic 
field (i.e., $\mathbf{B}=\mathbf{%
\nabla }\times \mathbf{A}=B_{\circ }\,\hat{z}$). We have shown that the corresponding KG-equation reduces to the two-dimensional radial
Schr\"{o}dinger Coulomb-like model (hence, the KG-Coulombic particle notion is used in the process). The exact textbook solution of which is used (along with the RG-modified energy-momentum dispersion relation (\ref{e2})) and the effects of rainbow gravity on the spectra are discussed.  We have also explored the effects of PDM (metaphorically speaking) on KG-Coulombic particles in cosmic string rainbow gravity and uniform magnetic field. In the process, we have studied the effects of four pairs of rainbow functions: (i) $g_{_{0}}\left(
y\right) =1$, $g_{_{1}}\left( y\right) =\sqrt{1-\epsilon y^{2}}$, (ii) $%
g_{_{0}}\left( y\right) =1$, $g_{_{1}}\left( y\right) =\sqrt{1-\epsilon y}$, (iii) $g_{_{0}}\left(
y\right) =g_{_{1}}\left( y\right) =\left( 1-\epsilon y\right) ^{-1}$, and (iv)  $g_{_{0}}( y) =( e^{\epsilon y}-1) /\epsilon y$  and $%
g_{_{1}}\left( y\right) =1$. 

Among the four pairs of rainbow functions, only $[g_{_{0}}\left(
y\right) =1$, $g_{_{1}}\left( y\right) =\sqrt{1-\epsilon y^{2}}]$ provided energy levels that are symmetric about $E=0$ line. Yet, it is interesting to observe that for this particular pair of rainbow functions, the energy levels are destined to be within the range $\sqrt{( k_{z}^{2}+m^{2}) /( 1+\delta
k_{z}^{2}) }\leq |E_{n_{r},\ell }|\,\leq \sqrt{1/\delta }=E_{p}/\sqrt{%
\epsilon }$ as the value of $|eB_\circ|$ increases from zero (i.e., zero charge $e=0$ and/or zero magnetic field strength, $B_\circ=0$). This effect is also observed for PDM (using the same pair of rainbow functions). This effect is documented in Figures 1(b) and 5(c). Evidently, moreover, for $\epsilon=1$ we obtain the maximum possible value of the energy,  $|E_{n_r,\ell}|_{max}$ ,  of the probe KG-Coulombic particle (in cosmic string rainbow gravity spacetime and a uniform magnetic field) as the Planck energy $E_p$ so that $\sqrt{( k_{z}^{2}+m^{2}) /( 1+
k_{z}^{2}/E_p^2) }\leq |E_{n_{r},\ell }|\,\leq \sqrt{1/\delta }=E_{p}$. 

We have, however, observed that the magnetic field did not remove the degeneracies of the energy levels associated with the magnetic quantum number $\ell=\pm |\ell|$, but the introduction of PDM-settings (through the PDM parameter $\eta$) has allowed the magnetic field to split $\ell=+|\ell|$ from $\ell=-|\ell|$.  Which is, in fact, a common feature for the four pairs of rainbow functions we have considered. Yet, without the PDM parameter $\eta$, we have noticed that all $S$-states (i.e., $\ell =0$ states) are degenerate with each other (positive with positive and negative with negative states) and have the same $%
K_{n_{r},0}$ value as shown in Eq. (\ref{e21-1}). However, when $\eta$ is brought into action, such degeneracy is removed (documented in (\ref{e33}) and Figures 5, 6, 7, and 8).

The current study, in fact, supports and emphasises Bezerra's et al. \cite{R81} statement that rainbow gravity is not just merely a mathematical re-scaling of both time and spatial coordinates. Rainbow gravity has deeply affected the spectroscopic structures for different rainbow function structures. The most interesting effect of which is observed  for the pair $[g_{_{0}}\left(
y\right) =1$, $g_{_{1}}\left( y\right) =\sqrt{1-\epsilon y^{2}}]$, which, in turn, implied that the energy of the probe KG-particle/antiparticle can not be more than the Planck's energy $E_p$. This result clearly suggests that the Planck energy $E_{p}$, in the rainbow gravity model, is not only yet another invariant energy scale alongside the speed
of light, but also a maximum possible  particle/antiparticle (here, KG-particles) energy value (e.g., \cite{R44}). More investigations should be carried out in this direction, we believe. Finally, to the best of our knowledge, the current methodical proposal did not appear elsewhere.


\begin{thebibliography}{99}
\bibitem{R1} J. Magueijo, L Smolin, Phys. Rev. Lett. \textbf{88} (2002)
190403.

\bibitem{R2} P. Galan, G. A. Mena Marugan, Phys. Rev. D \textbf{70} (2004)
124003.

\bibitem{R3} G. Amelino-Camelia, Int. J. Mod. Phys. D \textbf{11} (2002) 35.

\bibitem{R4} G. Amelino-Camelia, Int. J. Mod. Phys. D \textbf{11} (2002)
1643.

\bibitem{R5} .H. Hosseinpour, H. Hassanabadi, J. K\v{r}\'{\i}\v{z}, S.
Hassanabadi. B. C. L\"{u}tf\"{u}o\^{g}lu, Int. J. Geom. Methods Mod. Phys. 
\textbf{18} (2021) 2150224.

\bibitem{R6} G. Amelino-Camelia, J. R. Ellis, N. Mavromatos, D. V.
Nanopoulos, S. Sakar, Nature \textbf{393} (1998) 763.

\bibitem{R7} J. Magueijo, L. Smolin, Class. Quant. Gravit. \textbf{21}
(2004) 1725.

\bibitem{R8} V. B. Bezerra, H. F. Mota, C. R. Muniz, Eur. Phys. Lett. 
\textbf{120} (2017) 10005.
\bibitem{R81}V. B. Bezerra, I. P. Lobo, H. F. Mota, C. R. Muniz, Ann. Phys. \textbf{401} (2019)
162.
\bibitem{R9} L Smolin, Nucl. Phys. B \textbf{742} (2006) 142

\bibitem{R10} Y. Ling, X. Li, H. B. Zhang, Mod. Phys. Lett. A \textbf{22}
(2007) 2749.

\bibitem{R11} K. Sogut, M. Salti, O. Aydogdu, Ann. Phys. \textbf{431} (2021)
168556.

\bibitem{R12} E. E. Kangal, M Salti, O Aydogdu, K. Sogut, Phys. Scr. \textbf{%
96} (2021) 095301.

\bibitem{R13} J. Magueijo, L Smolin, Phys. Rev. D \textbf{67} (2003) 044017.

\bibitem{R14} M. Takeda et al, Astrophys. J. \textbf{522} (1999) 225.

\bibitem{R15} M. Takeda et al, Phys. Rev. Lett. \textbf{81} (1998) 1163.

\bibitem{R16} D. Finkbeiner, M. Davis, D. Schleged, Astrophys. J. \textbf{544%
} (2000) 81.

\bibitem{R17} D. Sudarsky, L. Urrutia, H. Vucetich, Phys. Rev. Lett. \textbf{%
89} (2002) 231301

\bibitem{R18} S. H. Hendi, M. Faizal, Phys. Rev. D \textbf{92} (2015) 044027.

\bibitem{R19} S. H. Hendi, Gen. Rel. Grav. \textbf{48} (2016) 50.

\bibitem{R20} S. H. Hendi, M. Faizal, B. Eslam Panah, S. Panahiyan, Eur.
Phys. J. C \textbf{76} (2016) 296.

\bibitem{R21} S. H. Hendi, S. Panahiyan, B. Eslam Panah, M. Momennia, Eur.
Phys. J. C \textbf{76} (2016) 150.

\bibitem{R211} B. Hamil, B. C. L\"{u}tf\"{u}o\^{g}lu, Int. J. Geom. Methods
Mod. Phys. \textbf{19} (2022) 2250047.

\bibitem{R22} S. H. Hendi, G. H. Bordbar, B. Eslam Panah, S. Panahiyan, J.
Cosmol. Astropart. Phys. \textbf{09} (2016) 013.

\bibitem{R221} Y. W. Kim, S. K. Kim, Y. J. Park, Eur. Phys. J C \textbf{76}
(2016) 557.

\bibitem{R222} S. H. Hendi, B. H. Panah,S. Panahiyan, Phys. Lett. B \textbf{%
769} (2017) 191.

\bibitem{R223} B. Panah, Phys. Lett. B \textbf{787} (2018) 45.

\bibitem{R224} R. Garattini, J. Cosmol. Astropart. Phys. \textbf{06} (2013)
017.

\bibitem{R23} M. Khodadi, K. Nozari, H. R. Sepangi, Gen. Rel. Grav. \textbf{%
48} (2016) 166.

\bibitem{R24} R. Garattini, J. Phys. Conf. Ser. \textbf{942} (2017) 012011.

\bibitem{R25} S. H. Hendi, M. Momennia, B. Eslam Panah, S. Panahiyan, Phys.
Dark Univ. \textbf{16} (2017) 26.

\bibitem{R26} V. B. Bezerra, H. R. Christiansen, M. S. Cunha, C. R. Muniz,
Phys. Rev. D \textbf{96} (2017) 024018.

\bibitem{R27} H. Aounallah, B. Pourhassan, S. H. Hendi, M. Faizal, Eur.
Phys. J. C \textbf{82} (2022) 351.

\bibitem{R28} K. Bakke, H. Mota, Eur. Phys. J. Plus \textbf{133} (2018) 409.

\bibitem{R29} K. Bakke, H. Mota, Gen. Rel. Grav. \textbf{52} (2020) 97.

\bibitem{R291} O. Mustafa, arXiv:2301.05464: "PDM KG-oscillators in cosmic
string rainbow gravity spacetime in a non-uniform magnetic field".

\bibitem{R30} O. von Roos, Phys. Rev. \textbf{B 27 }(1983) 7547.

\bibitem{R31} O. Mustafa, Phys. Lett. \textbf{A 384} (2020) 126265.

\bibitem{R32} O. Mustafa, S. H. Mazharimousavi, Int. J. Theor. Phys \ 
\textbf{46} (2007) 1786.

\bibitem{R33} O. Mustafa, Z. Algadhi, Eur. Phys. J. Plus \textbf{134} (2019)
228.

\bibitem{R34} M. A. F. dos Santos, I. S. Gomez, B. G. da Costa, O. Mustafa,
Eur. Phys. J. Plus \textbf{136 }(2021) 96.

\bibitem{R35} A. Khlevniuk, V. Tymchyshyn, J. Math. Phys. \textbf{59} (2018)
082901.

\bibitem{R36} O. Mustafa, J. Phys. \textbf{A}; Math. Theor. \textbf{48}
(2015) 225206.

\bibitem{R37} M. A. F. dos Santos, I. S. Gomez, B. G. da Costa, O. Mustafa,
Eur. Phys. J. Plus \textbf{136 }(2021) 96.

\bibitem{R38} O. Mustafa, Ann. Phys. \textbf{440} (2022) 168857.

\bibitem{R39} O. Mustafa, Eur. Phys. J. C \textbf{82} (2022) 82.

\bibitem{R40} O. Mustafa, Ann. Phys. \textbf{446} (2022) 169124.
\bibitem{R401} O. Mustafa, Phys. Scr. \textbf{98} (2023) 015302.
\bibitem{R402} O. Mustafa, Eur. Phys. J. Plus \textbf{138} (2023) 21.

\bibitem{R41} G. Amelino-Camelia, J. R. Ellis, N. Mavromatos, D. V.
Nanopoulos, Int. J. Mod. Phys. A \textbf{12} (1997) 607.

\bibitem{R42} G. Amelino-Camelia, Living Rev. Relativ. \textbf{16} (2013) 5.

\bibitem{R43} J. Magueijo, L. Smolin, Phys. Rev. Lett. \textbf{88} (2002)
190403.
\bibitem{R44} J. Hu, H. Hu, Results in Physics \textbf{43} (2022) 106082.
\end{thebibliography}
\end{document}